\documentclass{article}

\usepackage[preprint]{icml2026}
\usepackage{amsmath,amssymb,amsfonts}
\usepackage{graphicx}
\usepackage{booktabs}
\usepackage{microtype}
\usepackage{natbib}
\usepackage{url}
\usepackage{nicefrac}
\usepackage{xcolor}
\usepackage{multirow}
\usepackage{fontawesome5}
\usepackage[most]{tcolorbox}
\usepackage{setspace}
\usepackage{listings}
\usepackage{float}
\usepackage{tabularx}
\usepackage{longtable}
\usepackage{multicol}
\usepackage{algorithm}
\usepackage{algpseudocode}
\usepackage{placeins}
\usepackage{cuted}
\usepackage{hyperref}
\definecolor{mydarkblue}{rgb}{0,0.08,0.45}
\hypersetup{colorlinks=true, linkcolor=mydarkblue, citecolor=mydarkblue, filecolor=mydarkblue, urlcolor=mydarkblue}

\graphicspath{{figures/}{./}}
\newtcolorbox{promptbox}[1]{
  colback=gray!2!white,
  colframe=gray!80!black,
  fonttitle=\bfseries\large,
  title=#1,
  arc=2mm,
  boxrule=0.8pt,
  width=\textwidth,
  left=10pt,
  right=10pt,
  top=10pt,
  bottom=10pt,
  fontupper=\normalsize,
  before skip=15pt,
  after skip=15pt,
  breakable
}

\lstdefinestyle{agentcode}{
  language=Python,
  basicstyle=\ttfamily\footnotesize,
  backgroundcolor=\color{white},
  keywordstyle=\color{blue!60!black}\bfseries,
  commentstyle=\color{green!40!black}\itshape,
  stringstyle=\color{orange!70!black},
  showstringspaces=false,
  breaklines=true,
  frame=l,
  rulecolor=\color{teal!50},
  xleftmargin=1em,
  aboveskip=0.5em,
  belowskip=0.5em
}

\icmltitlerunning{DARE: Aligning LLM Agents with the R Statistical Ecosystem via Distribution-Aware Retrieval}

\begin{document}

\twocolumn[
\icmlsetsymbol{equal}{*}
\icmltitle{DARE: Aligning LLM Agents with the R Statistical Ecosystem via Distribution-Aware Retrieval}

\begin{icmlauthorlist}
  \icmlauthor{Maojun Sun}{equal,a}
  \icmlauthor{Yue Wu}{equal,a}
  \icmlauthor{Yifei Xie}{equal,a}
  \icmlauthor{Ruijian Han}{a}
  \icmlauthor{Binyan Jiang}{a}
  \icmlauthor{Defeng Sun}{b}
  \icmlauthor{Yancheng Yuan}{b}
  \icmlauthor{Jian Huang}{a,b}
\end{icmlauthorlist}

\icmlaffiliation{a}{Department of Data Science and Artificial Intelligence, The Hong Kong Polytechnic University, Hong Kong SAR, China}
\icmlaffiliation{b}{Department of Applied Mathematics, The Hong Kong Polytechnic University, Hong Kong SAR, China}

\icmlcorrespondingauthor{Ruijian Han}{ruijian.han@polyu.edu.hk}
\icmlcorrespondingauthor{Yancheng Yuan}{yancheng.yuan@polyu.edu.hk}
\icmlcorrespondingauthor{Jian Huang}{j.huang@polyu.edu.hk}

\begin{center}
  \small
  \faGlobe\ \href{https://ama-cmfai.github.io/DARE_webpage/}{Project Website}
  \qquad
  \href{https://huggingface.co/Stephen-SMJ/DARE-R-Retriever}{\includegraphics[height=1em]{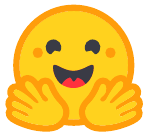}\ DARE-R-Retriever}
  \qquad
  \href{https://huggingface.co/datasets/Stephen-SMJ/RPKB}{\includegraphics[height=1em]{huggingface_logo.pdf}\ RPKB}
  \qquad
  \faGithub\ \href{https://github.com/AMA-CMFAI/DARE}{GitHub}
\end{center}

\icmlkeywords{Data Science Agents, Retrieval-Augmented Generation, Tool Learning, Representation Learning, R, Statistical Software}
\vskip 0.3in
]

\printAffiliationsAndNotice{\icmlEqualContribution}

\begin{abstract}
Large Language Model (LLM) agents excel at automating data science workflows, yet often underutilize rigorous statistical methods in R. Existing Retrieval-Augmented Generation (RAG) approaches rely on semantic similarity, ignoring the data-distribution and structural characteristics crucial for determining a statistical method's applicability. To address this, we propose DARE (Distribution-Aware Retrieval Embedding), a representation learning framework incorporating these data characteristics into R function retrieval. Our contributions are fourfold: (i) RPKB, a curated knowledge base of 8,191 high-quality R functions; (ii) DARE, the distribution-aware retrieval framework; (iii) RCodingAgent, an R-oriented agent integrating DARE for end-to-end analysis; and (iv) RStatBench, a comprehensive benchmark of 100 statistical tasks.
Experiments demonstrate that DARE improves NDCG@10 by 13.43–32.65 percentage points across evaluated embedding models, with a MiniLM-based DARE achieving 93.92\% NDCG@10 alongside high efficiency. Furthermore, integrating DARE into RCodingAgent nearly doubles mean downstream accuracy on RStatBench, increasing from 31.8\% to 62.2\% across ten LLMs. This work helps bridge modern LLM automation with R's established statistical computing ecosystem.

\end{abstract}

\begin{figure}[h]
  \centering
  \includegraphics[width=\linewidth]{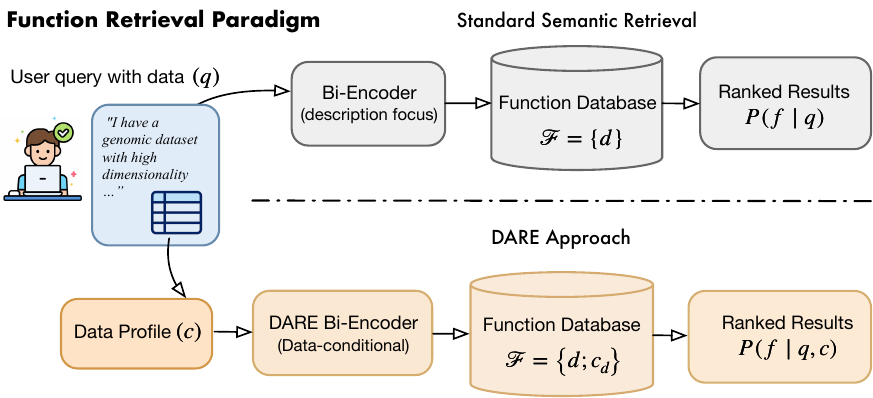}
  \caption{Comparison of traditional semantic function retrieval and the proposed Distribution-Aware Retrieval Embedding (DARE) method.\label{fig:com}}
\end{figure}

\section{Introduction}

\noindent Large Language Model (LLM)-based agents have rapidly emerged as a promising paradigm for automating data science workflows \citep{chen2025largelanguagemodelbaseddata, sun2026survey}. By integrating natural language reasoning, code generation, and tool execution, these agents are increasingly capable of translating user intent expressed in natural language into executable analytical pipelines. Recent systems have demonstrated strong performance in tasks such as data preprocessing, exploratory analysis, feature engineering, and model training \citep{sun2026dsaeval}, significantly reducing the manual effort traditionally required in end-to-end data science processes \citep{zhang2023data, sun2026lambda, hong2025data, zhang2025deepanalyze}.

R has long served as a core environment for statistical computing \citep{rcore}. The statistical community has developed a rich collection of theoretically grounded methods, including generalized linear modeling \citep{nelder1972generalized}, high-dimensional inference \citep{dezeure2015high}, survival analysis \citep{rai2021survival}, differential expression analysis in genomics \citep{Lin03042021}, and advanced time-series forecasting \citep{Burman01122006}. Many of these methods are implemented and maintained within the R ecosystem \citep{ihaka1996r, rcore}, particularly through the Comprehensive R Archive Network (\url{https://cran.r-project.org/}), which hosts thousands of specialized packages that encode domain expertise and methodological best practices.

However, this extensive statistical resource remains underutilized by current LLM-based data science agents, which predominantly operate within Python-centric ecosystems \citep{orlanski2023measuring, zhao2025current}. Because their training corpora are heavily skewed toward general-purpose programming languages, especially Python, these agents often exhibit substantial performance gaps when working with R. In practice, they may default to Python implementations even when more appropriate statistical methods are available in R, or generate R code that hallucinates function names, misconfigures parameters, and fails to identify the appropriate package.

A natural strategy to mitigate these limitations is RAG \citep{lewis2020retrieval}, where agents retrieve external documentation or function descriptions from R package repositories. Existing embedding models primarily rely on semantic similarity between user queries and textual descriptions of functions. However, the applicability of statistical methods often depends not only on semantic intent but also on data distribution characteristics, such as sparsity structure, dimensionality regimes, distributional assumptions, and modality-specific constraints. General-purpose embedding models, trained on broad web corpora, frequently fail to capture these subtle yet crucial distributional conditions, resulting in retrieval errors that propagate to downstream code generation and execution failures.

To address this limitation, we first construct RPKB (R Package Knowledge Base), a curated repository containing 8,191 high-quality R function records spanning diverse statistical domains. RPKB provides structured function metadata, documentation, and usage information, serving as a valuable resource for statistical tool retrieval and LLM tool learning.

Building upon RPKB, we propose DARE (Distribution-Aware Retrieval Embedding), a representation learning framework for distribution-aware R function retrieval. DARE augments function documentation with structured data profiles and fine-tunes an embedding backbone using symmetric contrastive learning, enabling the retriever to distinguish functions that are semantically similar but statistically incompatible under different data contexts. Across nine embedding models, DARE improves NDCG@10 by 13.43--32.65 percentage points over the corresponding off-the-shelf checkpoints. The DARE model, trained on all-MiniLM-L6-v2 \citep{reimers2019sentence}, achieves 93.92\% NDCG@10 with 23M parameters, 3.70 ms/query latency, and throughput of 8,359 queries/s on a single A100 GPU. It is designed as a lightweight and plug-and-play retrieval module that can be seamlessly integrated into LLM-based agent systems like LAMBDA \citep{sun2026lambda, sun2026rejoinder1}. Figure~\ref{fig:com} illustrates the difference between traditional semantic retrieval and DARE.

To investigate the practical impact of DARE in real-world statistical analysis workflows, we further design RCodingAgent, an R-oriented LLM agent that integrates DARE to conduct end-to-end statistical analyses through iterative code generation, execution, and refinement. We further construct RStatBench, a comprehensive benchmark of 100 R-based statistical analysis tasks spanning seven application domains and diverse data modalities, dimensionalities, and statistical methods. Across ten recent LLMs, integrating DARE into RCodingAgent increases mean downstream accuracy from 31.8\% to 62.2\% while reducing agent steps and token usage.







\section{Related Work}

\noindent \textbf{LLM Agents for Data Science.}\hspace{5pt} Recent works have empowered LLMs to act as data science agents by equipping them with planning, code execution environments, self-correction, and report generation \citep{sun2026survey,hong2025data, sun2026rejoinder2}. While these agents show promise, they predominantly rely on Python. Efforts to incorporate R are limited and often suffer from lower code generation quality due to the scarcity of R in training data relative to Python \citep{starcoder}.

\noindent \textbf{Dense Retrieval and RAG.}\hspace{5pt}
Dense retrieval maps queries and documents into a shared latent space using bi-encoders, as exemplified by DPR~\citep{karpukhin2020dense} and Contriever~\citep{izacard2021unsupervised}. Building on these methods, recent models such as \textit{Snowflake Arctic}~\citep{merrick2024embeddingclusteringdataimprove_snowflake_arctic}, \textit{GTE-large-en-v1.5}~\citep{gte-large-en-v1.5}, and \textit{MXBAI}~\citep{mxbai-large-v1} improve general retrieval through larger encoders and large-scale instruction tuning. However, statistical tool retrieval requires more than semantic similarity, because method applicability also depends on factors such as data distributions and modeling assumptions. Moreover, the larger size of these general-purpose encoders can increase query-encoding costs in interactive agent workflows. To address these limitations, DARE incorporates data-distribution information into representation learning, enabling accurate and efficient retrieval of statistically compatible R functions.

\noindent \textbf{Tool Learning.}\hspace{5pt} Tool learning aims to teach LLMs to use external APIs \citep{toolformer, shen2023hugginggpt}. Most existing approaches rely on in-context learning (ICL) with API descriptions, which tightly couples tool selection with generation. Our work differs by introducing a dedicated, data-aware retrieval module that decouples tool selection from generation, enabling scalable access to thousands of statistical functions. When augmented with DARE, LLM agents are able to effectively utilize a broader range of statistical packages and achieve substantial performance improvements in downstream data analysis tasks.

\section{Methodology}
In this section, we present details of the RPKB (R
Package Knowledge Base) construction, the proposed DARE framework, RCodingAgent, and RStatBench.
\begin{figure*}[t]
  \centering
  \includegraphics[width=\textwidth]{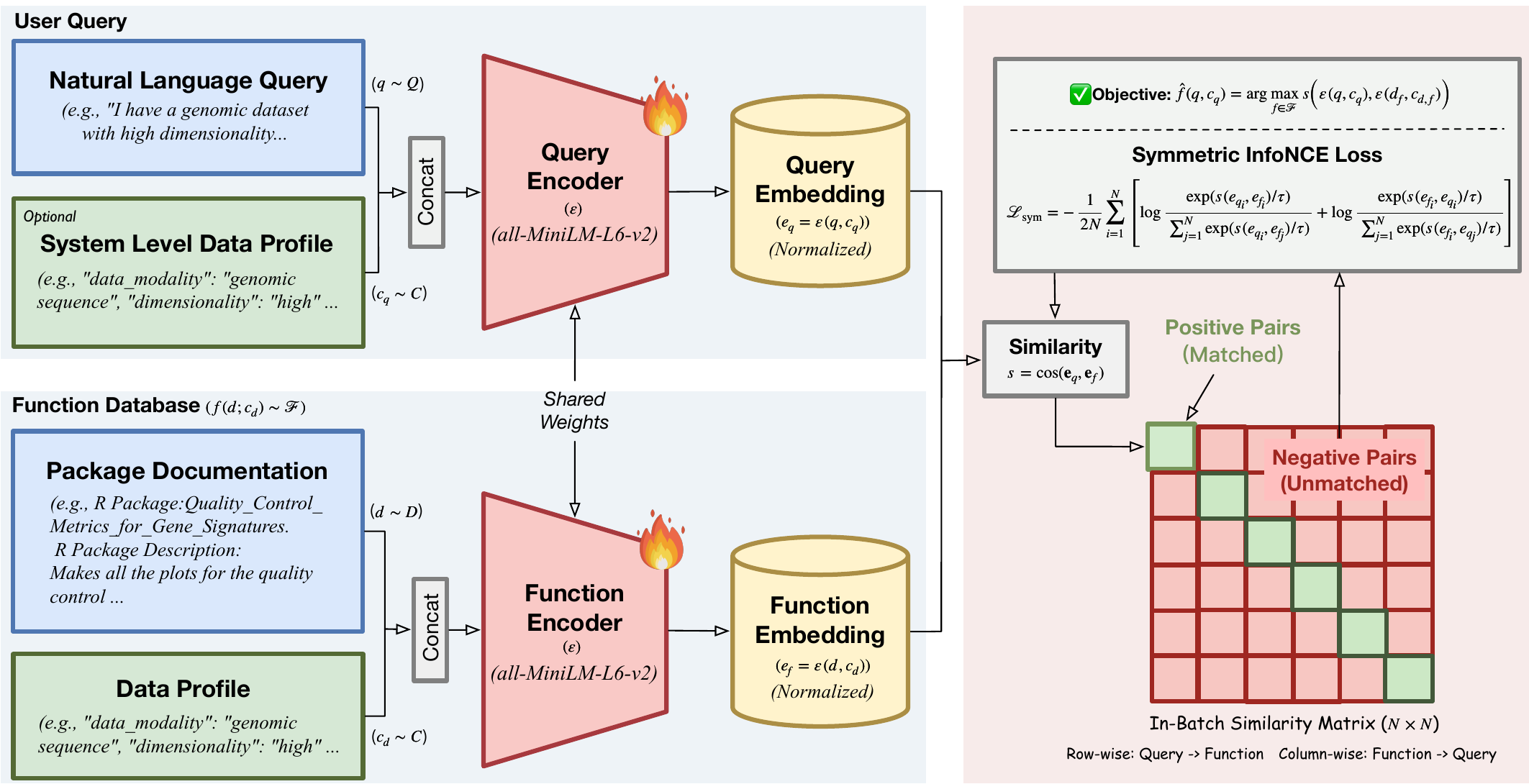}
  \caption{The overall framework of the DARE training process. \label{fig:dare}}
\end{figure*}
\subsection{RPKB Construction}
We constructed RPKB from the HTML and PDF documentation of 1,000 R packages on the Comprehensive R Archive Network (CRAN). We extracted package-level metadata (\textit{Package Name, Description, Version, Author}) and function-level documentation (\textit{Description, Usage, Arguments, Value}), then chunked the corpus by function. To retain statistically informative records, we excluded generic utilities such as I/O and basic string operations, functions with vague or incomplete descriptions, and auxiliary methods lacking specific algorithmic content. For each retained function, \textit{Grok-4.1-fast} inferred a data-requirements profile from its documentation, including data modality, distribution assumptions, and dimensionality. The resulting records were indexed using ChromaDB, yielding a retrieval corpus $\mathcal{F}$ of 8,191 R functions spanning diverse statistical domains. Further details on construction, selection, examples, and generation prompts are provided in the Supplementary Materials.

\subsection{DARE Framework}
Let $\mathcal{F}=\{f_k\}_{k=1}^{|\mathcal{F}|}$ denote the candidate R functions in RPKB. Each function is represented as $f=(d,c_d)$, where $d$ is its documentation and $c_d$ is its data-requirements profile. Given a user request $q$, DARE ranks functions according to their relevance to the analytical task and its data characteristics.

DARE uses a shared-weight bi-encoder initialized from a pretrained text embedding model, such as \textit{sentence-transformers/all-MiniLM-L6-v2}~\citep{reimers2019sentence}. The encoder $\varepsilon(\cdot)$ maps query and function texts into a common $m$-dimensional space, and $[\cdot\,;\,\cdot]$ denotes textual concatenation. For natural-language requests, the query embedding is $\mathbf{e}_q=\varepsilon(q)$, while the function embedding is $\mathbf{e}_f=\varepsilon([d;c_d])$. Training queries express data characteristics with varying levels of detail. A subset additionally follows the explicit $[q;c_q]$ format, with query embedding $\varepsilon([q;c_q])$. Here, $c_q$ represents the kind of data information that RCodingAgent can extract from an uploaded dataset when the request itself lacks sufficient detail (Section~\ref{sec:agent}).

We compute relevance by cosine similarity, $s(\mathbf{e}_q,\mathbf{e}_f)=\mathbf{e}_q^{\top}\mathbf{e}_f/(\lVert\mathbf{e}_q\rVert_2\lVert\mathbf{e}_f\rVert_2)$, and retrieve $\hat f=\arg\max_{f\in\mathcal F}s(\mathbf{e}_q,\mathbf{e}_f)$. After normalization, retrieval can be implemented using Maximum Inner Product Search (MIPS) over precomputed function embeddings $\{\mathbf{e}_f\}_{f\in\mathcal F}$.

We construct a supervised retrieval dataset $\mathcal{D}=\{(q_i, c_{q,i}, f_i)\}_{i=1}^M$, where $f_i$ denotes the ground-truth function (comprising documentation $d_i$ and profile $c_{d,i}$) for the $i$-th query-context pair.
Each query $q_i$ is generated from the documentation describing the functionality of its target function $f_i$. In this study, we consider one target function per query.

The framework uses a symmetric InfoNCE objective with in-batch negatives \citep{oord2019representationlearningcontrastivepredictive}. For a mini-batch of $N$ query--function pairs, each $f_i$ is the positive function for $q_i$; the other functions and queries in the batch serve as negatives in the two respective directions. With learnable temperature $\tau$, let $z_{ij}=s(\mathbf{e}_{q_i},\mathbf{e}_{f_j})/\tau$. The objective is
\begingroup
\setlength{\abovedisplayskip}{4pt}
\setlength{\belowdisplayskip}{4pt}
\setlength{\abovedisplayshortskip}{0pt}
\setlength{\belowdisplayshortskip}{3pt}
\setlength{\jot}{1pt}
\begin{equation}
\label{eq:symmetric_infonce}
\begin{aligned}
\mathcal{L}_{\mathrm{sym}}
&=\frac{1}{2N}\sum_{i=1}^{N}
\left(\log\sum_{j=1}^{N}e^{z_{ij}}-z_{ii}\right)\\
&\quad+\frac{1}{2N}\sum_{i=1}^{N}
\left(\log\sum_{j=1}^{N}e^{z_{ji}}-z_{ii}\right).
\end{aligned}
\end{equation}
\endgroup
In Equation~\eqref{eq:symmetric_infonce}, the first term identifies the paired function for each query, whereas the second identifies the paired query for each function. This objective maximizes the similarity between each query embedding $\mathbf{e}_{q_i}$ and its corresponding function embedding $\mathbf{e}_{f_i}$, while minimizing its similarity to other candidate function embeddings $\mathbf{e}_{f_j}$ that do not align with the query's semantic and distributional constraints. The symmetric term applies the same contrast in the function-to-query direction. Figure~\ref{fig:dare} illustrates the framework of DARE.

\subsection{RCodingAgent: DARE-Augmented Agentic Data Analysis}\label{sec:agent}

\begin{figure*}[tbp]
  \centering
    \includegraphics[width=\textwidth]{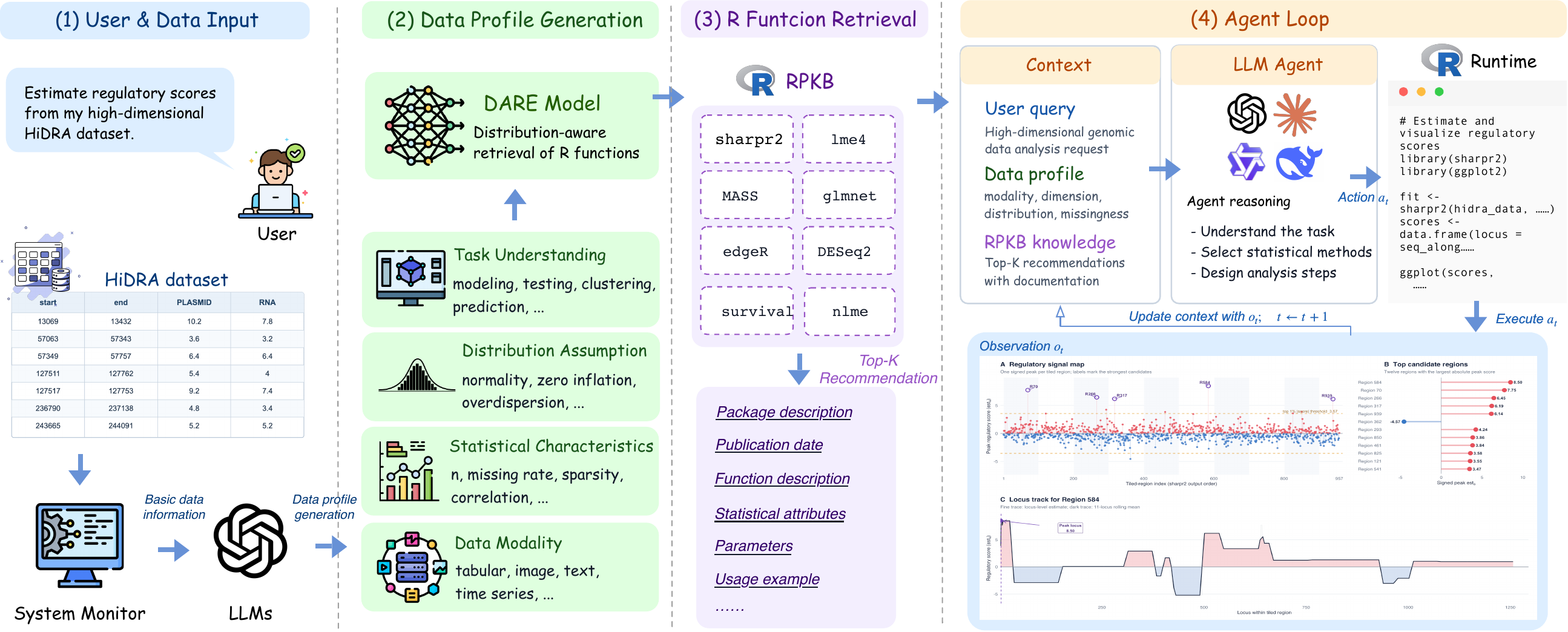}
  \caption{Framework of RCodingAgent. Given a user query and dataset, the system generates a data profile, retrieves compatible R functions from RPKB using DARE, and iteratively generates and executes R code in the Agent Loop. \label{fig:dare_agent}}
\end{figure*}

We further design RCodingAgent, an LLM-based agent that incorporates the DARE model to automate statistical analysis end to end. As illustrated in Figure~\ref{fig:dare_agent}, the system receives the user's request $q$ and, when available, a dataset $c$. It inspects the dataset to obtain data information $c_q$ relevant to statistical method selection. This information can be appended to the request as $[q;c_q]$ or used by an LLM to rewrite the request. DARE uses the resulting query to retrieve the top three R functions from RPKB and provides their documentation to the agent. The agent then reasons about the task, plans the analysis, generates R code, and executes it in a sandbox. Each execution returns an observation that informs the agent's next step, allowing it to refine or advance the analysis until it produces a final answer. The algorithm for this loop and a representative execution example are provided in Supplementary Material Section E.

\subsection{RStatBench: Benchmarking LLM Agents on R Statistical Analysis Tasks}

\begin{figure*}[tbp]
  \centering
  \includegraphics[width=\textwidth]{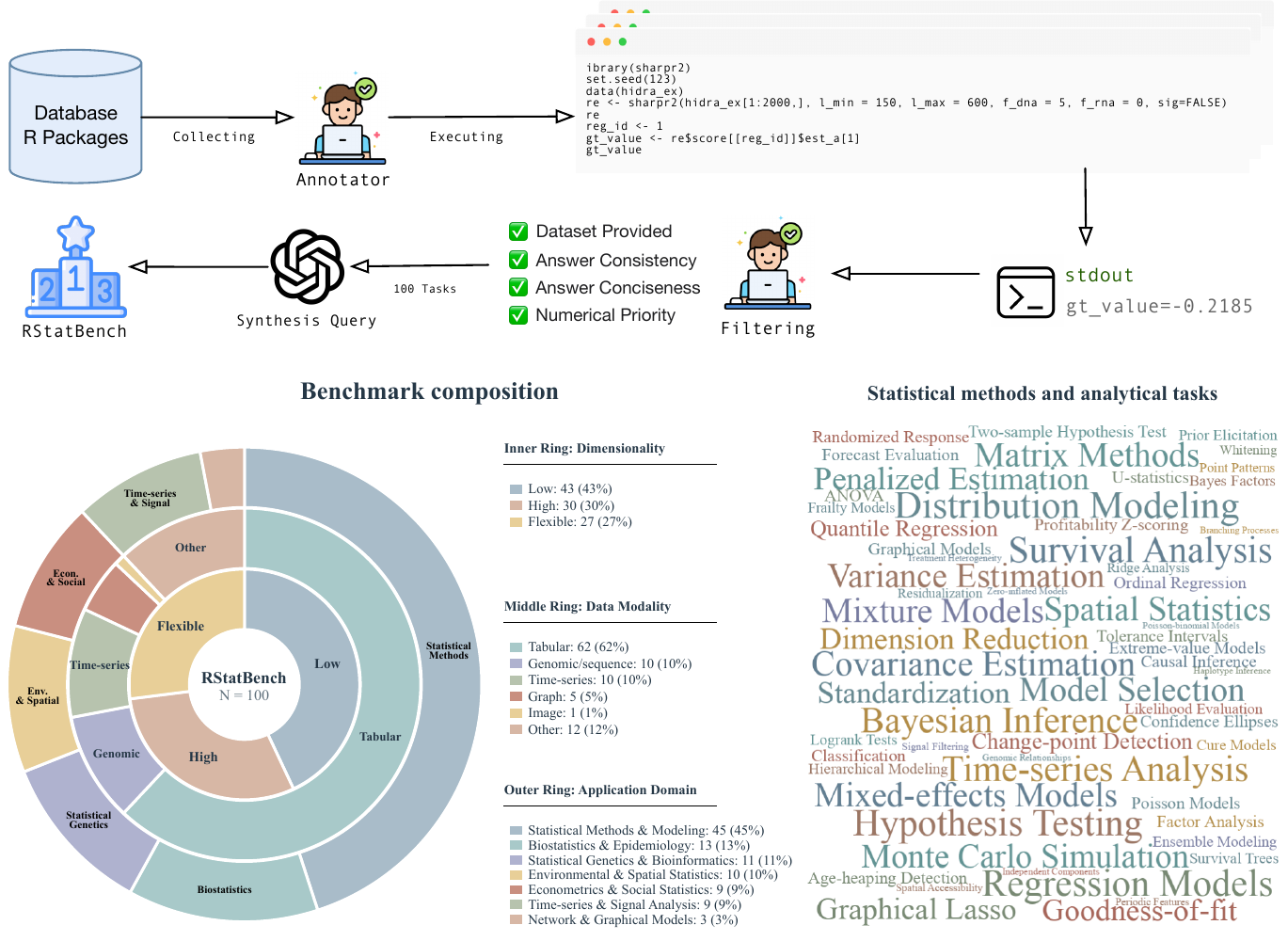}
  \caption{Composition of RStatBench and a word cloud showing the frequency of tasks in the benchmark.}
  \label{fig:rstatbench}
\end{figure*}

To systematically evaluate LLM agents in realistic, end-to-end R-oriented statistical programming workflows, we design RStatBench, a suite of 100 representative R-based statistical analysis tasks. Specifically, we first collect high-quality example R scripts from the RPKB repository covering diverse statistical domains, then execute these scripts to produce verified ground-truth outputs. Based on the underlying datasets and analytical objectives, we prompt LLMs to generate natural user-style
analytical queries. Each generated query is manually verified and paired with its corresponding ground-truth output to ensure correctness and reproducibility. Detailed task specifications and query generation prompts are provided in the Supplementary Materials.

Figure~\ref{fig:rstatbench} summarizes the composition of RStatBench. The benchmark covers a diverse range of statistical settings. The tasks span domains of statistical methods and modeling, biostatistics and epidemiology, statistical genetics and bioinformatics, environmental and spatial statistics, econometrics and social statistics, time-series and signal analysis, and network and graphical models. The benchmark also incorporates multiple data modalities, including tabular, genomic or sequence, time-series, graph, and image data, as well as tasks with different dimensionality characteristics. The word cloud on the right further illustrates the breadth of statistical methods and analytical objectives represented in the benchmark, ranging from classical techniques such as hypothesis testing, ANOVA, regression, and survival analysis to more specialized topics such as graphical lasso, change-point detection, mixed-effects models, Monte Carlo simulation, and causal inference.

\section{Experiments}

\subsection{Experimental Setup}
We organize the experiments around five research questions concerning DARE and its application to LLM-based statistical analysis.

\noindent\textbf{RQ1: Does DARE effectively improve R-function retrieval across embedding models?}
We use \textit{MiMo-V2-flash} to generate 30 queries per source function with five prompt templates ranging from simple to complex, reflecting different levels of data-analysis expertise. The dataset contains 245,738 queries, each associated with one source function as ground truth, comprising 212,974 training queries and 32,764 test queries under a query-level split. We use DARE to train nine models, including \textit{sentence-transformers/all-MiniLM-L6-v2} \citep{reimers2019sentence}, \textit{jinaai/jina-embeddings-v2-base-en} \citep{jinaai-jina-embedding}, \textit{Alibaba-NLP/gte-large-en-v1.5} \citep{gte-large-en-v1.5}, \textit{sentence-transformers/all-mpnet-base-v2} \citep{all-mpnet-base-v2}, \textit{WhereIsAI/UAE-Large-V1} \citep{uae-large-v1}, \textit{mixedbread-ai/mxbai-embed-large-v1} \citep{mxbai-large-v1}, \textit{BAAI/bge-m3} \citep{bge-m3embedding}, \textit{Snowflake/arctic-embed-l} \citep{merrick2024embeddingclusteringdataimprove_snowflake_arctic}, and \textit{intfloat/e5-large-v2} \citep{intfloat-eg-large}. For each model, we compare its off-the-shelf checkpoint with the model fine-tuned using DARE. All backbones are trained for 100 epochs with symmetric InfoNCE, AdamW \citep{adamw}, and batch size 256. Detailed results are shown in Table~\ref{tab:matched_backbone_finetuning}. Beyond the query-level split, we further study function-level and package-level splits, which hold out the target functions or their entire packages from training, and held-out-style splits, which exclude some query styles from training. These settings test out-of-distribution (OOD) generalization beyond the functions, packages, or query styles seen during training.

\noindent\textbf{RQ2: Which fine-tuned backbone offers the best accuracy--efficiency trade-off?}
We compare retrieval quality across the nine fine-tuned backbones alongside query-encoding latency and throughput. Both efficiency measures are obtained on a single NVIDIA A100 GPU in FP16, using batch size 1 for latency and batch size 128 for throughput. This allows us to compare retrieval performance with the computational cost of encoding queries. Results are shown in Figure~\ref{fig:qps}.

\noindent\textbf{RQ3: Does DARE learn to perceive data profiles?}
To identify the sources of DARE's retrieval gains, we conduct ablations that vary fine-tuning, function-side data profiles, and query-side data information. The first three comparisons use the full 32,764-query test set; a separate comparison evaluates 1,763 queries with explicit query-side data information. To examine whether retrieval depends on lexical overlap, we select 9,331 test queries containing distributional or structural terms and compare each query with a version using alternative expressions, such as replacing ``Gaussian'' with ``bell-shaped'' and ``sparse'' with ``mostly-zero.'' We further test statistical compatibility across different distributions, tasks, and conflicting query and data information. Results are reported in Tables~\ref{tab:r1_matched_profile_contrasts} and~\ref{tab:r1_compatibility_stress}.

\noindent\textbf{RQ4: Does DARE retrieval improve end-to-end statistical analysis?}
We evaluate RCodingAgent with ten LLMs on the 100 RStatBench tasks, with DARE retrieval enabled and disabled. The agent configuration, R environment, and 20-step limit are identical in both conditions. We measure the effect of retrieved R-function guidance on task accuracy, wall-clock duration, agent steps, token usage, and cost. Supplementary Material Section~L provides model, run, and pricing details. Figures~\ref{fig:overall} and~\ref{fig:agent_efficiency} summarize the results.

\noindent\textbf{RQ5: How does performance vary across statistical domains, and which errors persist?}
We further evaluate RCodingAgent across application domains, including biostatistics and epidemiology, statistical genetics and bioinformatics, and econometrics and social statistics. We also examine agent traces to identify recurring errors. Results are presented in Subsection~\ref{sec:error_analysis}.

\vspace{-6pt}
\subsection{Evaluation Metrics}
Let $\mathcal{Q}$ denote the set of evaluation queries with size $|\mathcal{Q}|$. For each query, we obtain a ranked list of retrieved functions $(f_{(1)},\ldots,f_{(K)})$.
Let $\mathrm{rel}(f)\in\{0,1\}$ indicate whether a retrieved function is matched.

\noindent \textbf{Recall@$k$.}\hspace{5pt}
For each query $q \in \mathcal{Q}$, we obtain a ranked list $\left(f_{(1)}^{(q)}, \ldots, f_{(K)}^{(q)}\right)$. Recall@$k$ is the fraction of queries whose ground-truth function appears in the top-$k$ results: $\mathrm{Recall}@k=|\mathcal{Q}|^{-1}\sum_{q\in\mathcal{Q}}\mathbb{I}\{\exists\,j\le k:\mathrm{rel}(f_{(j)}^{(q)})=1\}$. Recall@1 corresponds to top-1 accuracy.

\noindent \textbf{NDCG@$k$.}\hspace{5pt} To capture rank sensitivity, we also report NDCG@$k$.
The discounted cumulative gain is $\mathrm{DCG}@k=\sum_{j=1}^{k}[2^{\mathrm{rel}(f_{(j)})}-1]/\log_2(j+1)$, and $\mathrm{NDCG}@k=\mathrm{DCG}@k/\mathrm{IDCG}@k$, where $\mathrm{IDCG}@k$ is the maximum achievable DCG@k under an ideal ranking. With one labeled relevant function per query, $\mathrm{IDCG}@k=1$, and NDCG@k is averaged over queries. For each query, it reduces to a logarithmically discounted score of how early the correct tool is retrieved.

\noindent \textbf{Mean Reciprocal Rank (MRR@$k$).}\hspace{5pt}
MRR@$k$ evaluates how far down the ranked list the first relevant function appears: $\mathrm{MRR}@k=|\mathcal{Q}|^{-1}\sum_{q\in\mathcal{Q}}1/\mathrm{rank}^{(q)}$, where $\mathrm{rank}^{(q)}$ is the position of the first ground-truth function for query $q$. If no relevant function is retrieved within the top $k$ results, its reciprocal rank is set to $0$.

\noindent \textbf{Efficiency Metrics.}\hspace{5pt}
To evaluate the deployment feasibility of DARE in real-time agentic workflows, we report two key efficiency indicators:
\begin{itemize}
    \item \textbf{Average Latency ($L$):} Measures the mean wall-clock time required to encode a single query-context pair. Given a set of evaluation queries $\mathcal{Q}$ and total inference time $T_{seq}$ under sequential execution (batch size $=1$), $L=T_{seq}|\mathcal{Q}|^{-1}\times1000$ ms/query.

    \item \textbf{Throughput (QPS):} Measures processing capacity under parallel execution. Given $|\mathcal{Q}|$ queries processed in optimal batches (e.g., batch size $=128$) with total inference time $T_{batch}$, $\mathrm{QPS}=|\mathcal{Q}|/T_{batch}$ queries/s.
\end{itemize}


\noindent \textbf{Success Rate (SR).}\hspace{4pt} To evaluate end-to-end agent performance, let $A(q)$ be the execution output for query $q$ and $G(q)$ the corresponding ground-truth answer. We define $\mathbb{I}(q)=1$ if $A(q)=G(q)$ and $0$ otherwise, and report $\mathrm{SR}=|\mathcal{Q}|^{-1}\sum_{q\in\mathcal{Q}}\mathbb{I}(q)$.

\section{Experimental Results}

We present the experimental results in the order of the five research questions in Section~4.1.

\subsection{RQ1: Retrieval Performance Across Embedding Models}
Table~\ref{tab:matched_backbone_finetuning} shows the performance of the nine embedding backbones before and after DARE fine-tuning. Despite substantial differences in their off-the-shelf performance, all backbones consistently improve on NDCG@10, MRR@10, Recall@10, and Recall@1. The NDCG@10 gain ranges from 13.43 percentage points for Snowflake Arctic Embed L to 32.65 points for MiniLM; after fine-tuning, the nine models achieve 91.53--94.40\% NDCG@10. Jina and GTE obtain the highest NDCG@10 values, at 94.40\% and 94.36\%, respectively, while Recall@10 exceeds 98\% for every backbone. Thus, the gains extend beyond a single ranking metric: the correct function is both more likely to be retrieved within the top ten and ranked nearer the top of that list. Given its much smaller size and comparable retrieval performance, we use MiniLM as the backbone for all subsequent experiments.

\begin{table*}[tbp]
\begin{singlespace}
\centering
\caption{Retrieval performance before and after DARE fine-tuning. Values in parentheses are the corresponding results before fine-tuning.}
\label{tab:matched_backbone_finetuning}
\small
\resizebox{\textwidth}{!}{%
\begin{tabular}{lcrrrr}
\toprule
Models & Parameters & NDCG@10 (\%) & MRR@10 (\%) & Recall@10 (\%) & Recall@1 (\%) \\
\midrule
Jina Embeddings v2 Base EN & 137 M & \textbf{94.40} ($\uparrow$ 74.29) & \textbf{93.00} ($\uparrow$ 69.65) & 98.61 ($\uparrow$ 88.73) & 89.39 ($\uparrow$ 59.69) \\
GTE Large EN v1.5 & 434 M & 94.36 ($\uparrow$ 66.39) & 92.98 ($\uparrow$ 61.22) & 98.53 ($\uparrow$ 82.57) & \textbf{89.42} ($\uparrow$ 50.40) \\
All-MPNet-Base-v2 & 110 M & 94.11 ($\uparrow$ 66.06) & 92.57 ($\uparrow$ 60.57) & \textbf{98.75} ($\uparrow$ 83.30) & 88.59 ($\uparrow$ 49.37) \\
UAE Large v1 & 335 M & 94.05 ($\uparrow$ 70.66) & 92.58 ($\uparrow$ 65.56) & 98.46 ($\uparrow$ 86.58) & 88.81 ($\uparrow$ 54.79) \\
MXBAI Embed Large v1 & 335 M & 93.92 ($\uparrow$ 70.68) & 92.45 ($\uparrow$ 65.65) & 98.38 ($\uparrow$ 86.39) & 88.64 ($\uparrow$ 55.08) \\
BGE-M3 & 568 M & 93.83 ($\uparrow$ 73.08) & 92.30 ($\uparrow$ 68.43) & 98.45 ($\uparrow$ 87.58) & 88.37 ($\uparrow$ 58.47) \\
Snowflake Arctic Embed L & 335 M & 92.75 ($\uparrow$ 79.32) & 90.91 ($\uparrow$ 75.10) & 98.31 ($\uparrow$ 92.35) & 86.12 ($\uparrow$ 65.49) \\
E5 Large v2 & 335 M & 91.53 ($\uparrow$ 75.13) & 89.26 ($\uparrow$ 70.86) & 98.38 ($\uparrow$ 88.38) & 83.46 ($\uparrow$ 61.52) \\
MiniLM-L6-v2 & 23 M & 93.92 ($\uparrow$ 61.27) & 92.33 ($\uparrow$ 55.53) & 98.71 ($\uparrow$ 79.36) & 88.23 ($\uparrow$ 44.12) \\
\bottomrule
\end{tabular}}
\end{singlespace}
\end{table*}



In the data-split experiments, function- and package-level holdouts reduce MiniLM's NDCG@10 from 93.92\% to 85.08\% and 85.75\%, respectively, indicating a clear loss under these OOD conditions. Nevertheless, the 23 M-parameter MiniLM retriever remains above the strongest off-the-shelf baseline, the 335 M-parameter Snowflake Arctic Embed L (79.32\% NDCG@10). In contrast, held-out query styles show no comparable degradation, with NDCG@10 ranging from 92.34\% to 94.73\%. Training curves are reported in Supplementary Material Section D, and full split results in Section J.

\subsection{RQ2: Accuracy-efficiency Trade-off}

\begin{figure*}[tbp]
  \centering
  \includegraphics[width=\textwidth]{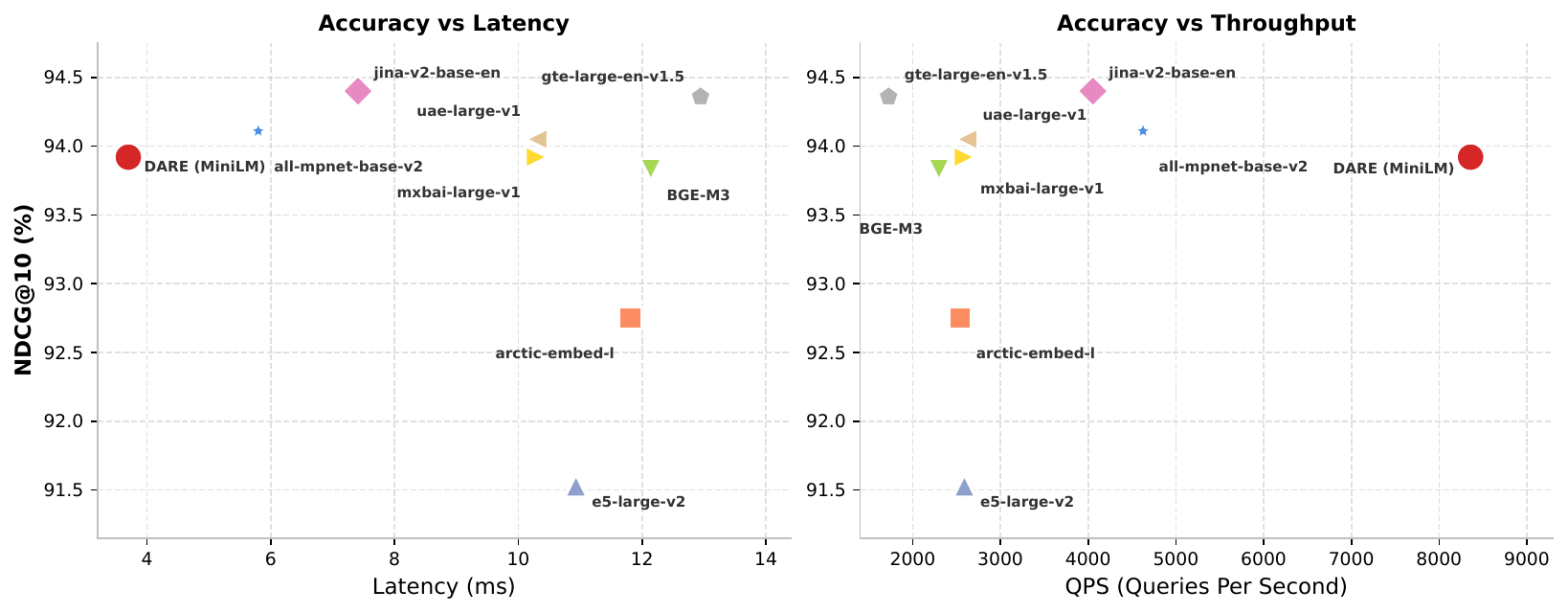}
  \caption{Accuracy--efficiency trade-off across fine-tuned backbones. The upper-left corner indicates a better accuracy--latency trade-off in the left panel, whereas the upper-right corner indicates a better accuracy--throughput trade-off in the right panel.}
  \label{fig:qps}
\end{figure*}

We evaluate each model's encoding efficiency alongside retrieval accuracy, measuring single-query latency and batched throughput in queries per second. As shown in Figure~\ref{fig:qps}, the MiniLM-based DARE model achieves 93.92\% NDCG@10 with a latency of 3.70 ms per query and a batched throughput of 8,359 queries/s. Jina achieves 94.40\% NDCG@10 but requires 7.41 ms per query and processes 4,055 queries/s; GTE achieves 94.36\% NDCG@10 at 12.95 ms per query and 1,725 queries/s. These models achieve slightly higher retrieval accuracy than MiniLM, but at the cost of substantially higher latency and lower throughput. MiniLM thus offers the best accuracy–efficiency trade-off among the evaluated backbones, and we use it for the subsequent experiments (MiniLM-L6-v2-based DARE model).

\subsection{RQ3: Does DARE Learn to Perceive Data Profiles?}
\label{sec:revision_controls}
Table~\ref{tab:r1_matched_profile_contrasts} presents an ablation study identifying the sources of DARE's retrieval gains. In this setting, $q$ denotes the user query, $d$ the function documentation, $c_d$ the documentation-derived function profile stored in RPKB, $q'$ a cleaned query from which explicit data-related information has been removed, and $q_s$ the query obtained by replacing distributional and structural terms in $q$ with alternative expressions.

\begin{table*}[tbp]
\centering
\small
\setlength{\tabcolsep}{3pt}
\caption{Ablation study of fine-tuning, function-side data profiles, query-side data information, and lexical overlap. Here, $q$ denotes the user's natural-language request, $c_q$ the system-generated data information, $q'$ a cleaned query with explicit data-related information removed, $d$ the function documentation, $c_d$ the documentation-derived function profile stored in RPKB, and $q_s$ the query obtained by replacing distributional and structural terms in $q$ with alternative expressions.}
\label{tab:r1_matched_profile_contrasts}
\begin{tabular*}{0.85\textwidth}{@{\extracolsep{\fill}}lcrrrrr}
\toprule
Setting & FT & \shortstack{Queries} & \shortstack{NDCG@10 (\%)} & \shortstack{MRR@10 (\%)} & \shortstack{Recall@10 (\%)} & \shortstack{Recall@1 (\%)} \\
\midrule
\multicolumn{7}{l}{\textbf{1. Contribution of fine-tuning}} \\
$q$ vs. $[d;c_d]$ & No & 32,764 & 61.27 & 55.53 & 79.36 & 44.12 \\
$q$ vs. $[d;c_d]$ & Yes & 32,764 & 93.92 & 92.33 & 98.71 & 88.23 \\
\midrule
\multicolumn{7}{l}{\textbf{2. Contribution of distributional features beyond fine-tuning}} \\
$q$ vs. $d$ & Yes & 32,764 & 84.89 & 82.46 & 92.49 & 77.25 \\
$q$ vs. $[d;c_d]$ & Yes & 32,764 & 93.92 & 92.33 & 98.71 & 88.23 \\
\midrule
\multicolumn{7}{l}{\textbf{3. Remove all data information}} \\
$q'$ vs. $d$ & Yes & 32,764 & 66.04 & 62.47 & 77.27 & 55.05 \\
$q$ vs. $d$ & Yes & 32,764 & 84.89 & 82.46 & 92.49 & 77.25 \\
\midrule
\multicolumn{7}{l}{\textbf{4. Contribution of fine-tuning with profiles on both sides}} \\
$[q;c_q]$ vs. $[d;c_d]$ & No & 1,763 & 57.41 & 51.99 & 74.42 & 41.07 \\
$[q;c_q]$ vs. $[d;c_d]$ & Yes & 1,763 & 96.62 & 95.91 & 98.75 & 93.93 \\
\midrule
\multicolumn{7}{l}{\textbf{5. Data-profile awareness or lexical overlap}} \\
$q$ vs. $[d;c_d]$ & Yes & 9,331 & 94.74 & 93.46 & 98.56 & 90.02 \\
$q_s$ vs. $[d;c_d]$ & Yes & 9,331 & 93.54 & 92.02 & 98.14 & 88.03 \\
\bottomrule
\end{tabular*}
\end{table*}

The results show that both fine-tuning and function-side profiles contribute to retrieval performance. Fine-tuning $q$ versus $[d;c_d]$ raises NDCG@10 from 61.27\% to 93.92\%. Among fine-tuned settings, adding $c_d$ to the documentation raises NDCG@10 from 84.89\% to 93.92\% and Recall@1 from 77.25\% to 88.23\%. The profile therefore provides a substantial gain beyond documentation-only fine-tuning. Query-side data descriptions also matter. Removing them from $q$ to obtain $q'$ reduces NDCG@10 from 84.89\% to 66.04\% and Recall@1 from 77.25\% to 55.05\% in the fine-tuned documentation-only setting. For example, ``I need to cluster samples in my high-dimensional, sparse gene-expression dataset, samples.csv'' may become ``I need to cluster samples in samples.csv.'' The task remains clear, but the data characteristics needed to distinguish among plausible clustering methods are lost.

In the idealized setting where the query is augmented by system-generated data information $c_q$, fine-tuning raises NDCG@10 from 57.41\% to 96.62\% and Recall@1 from 41.07\% to 93.93\% on the same 1,763-query subset. This gain indicates that DARE learns to align query-side data information with function-side requirements.  Finally, we test whether retrieval depends on lexical overlap between queries and function profiles. On the same 9,331 test queries, replacing distributional and structural terms with alternative expressions changes NDCG@10 from 94.74\% to 93.54\%. The similar performance suggests that DARE captures the distributional meaning conveyed by the query, supporting data-profile awareness rather than reliance solely on exact lexical overlap.

We further assess whether DARE captures statistical compatibility across distributional requirements and analytical tasks. As shown in Table~\ref{tab:r1_compatibility_stress}, DARE achieves 92.06\% NDCG@10 on 2,584 queries grouped by shared analytical intent but differing data distributions, indicating that its rankings remain sensitive to distributional differences. On 29,364 queries grouped by similar data profiles but differing tasks, it achieves 93.87\% NDCG@10, suggesting that profile similarity does not obscure the requested analytical goal. We also construct 112 cases in which the query's distributional description conflicts with the correct data information. Appending that information yields 89.97\% NDCG@10; rewriting the query with GPT-5.6-Luna to resolve the conflict raises the score to 96.64\%. Together with the terminology-substitution results, these findings suggest that DARE captures both distributional compatibility and analytical intent rather than relying solely on lexical overlap. Supplementary Material Section~K describes the test construction and an additional reference condition.

\begin{table*}[t]
\centering
\small
\caption{Retrieval performance across statistical-compatibility settings.}
\label{tab:r1_compatibility_stress}
\resizebox{\textwidth}{!}{%
\begin{tabular}{lrrrr}
\toprule
Setting & NDCG@10 (\%) & MRR@10 (\%) & Recall@10 (\%) & Recall@1 (\%) \\
\midrule
Same intent, different distribution & 92.06 & 90.05 & 98.14 & 84.91 \\
Same profile, different task & 93.87 & 92.27 & 98.69 & 88.17 \\
\midrule
Contradictory query with correct $c_q$ & 89.97 & 87.16 & 98.21 & 78.57 \\
LLM rewrite & 96.64 & 95.50 & 100.00 & 91.96 \\
\bottomrule
\end{tabular}}
\end{table*}

\subsection{RQ4: Downstream Agent Performance and Efficiency}

\begin{figure*}[t]
\centering
\includegraphics[width=\textwidth,trim=28bp 14bp 7bp 16bp,clip]{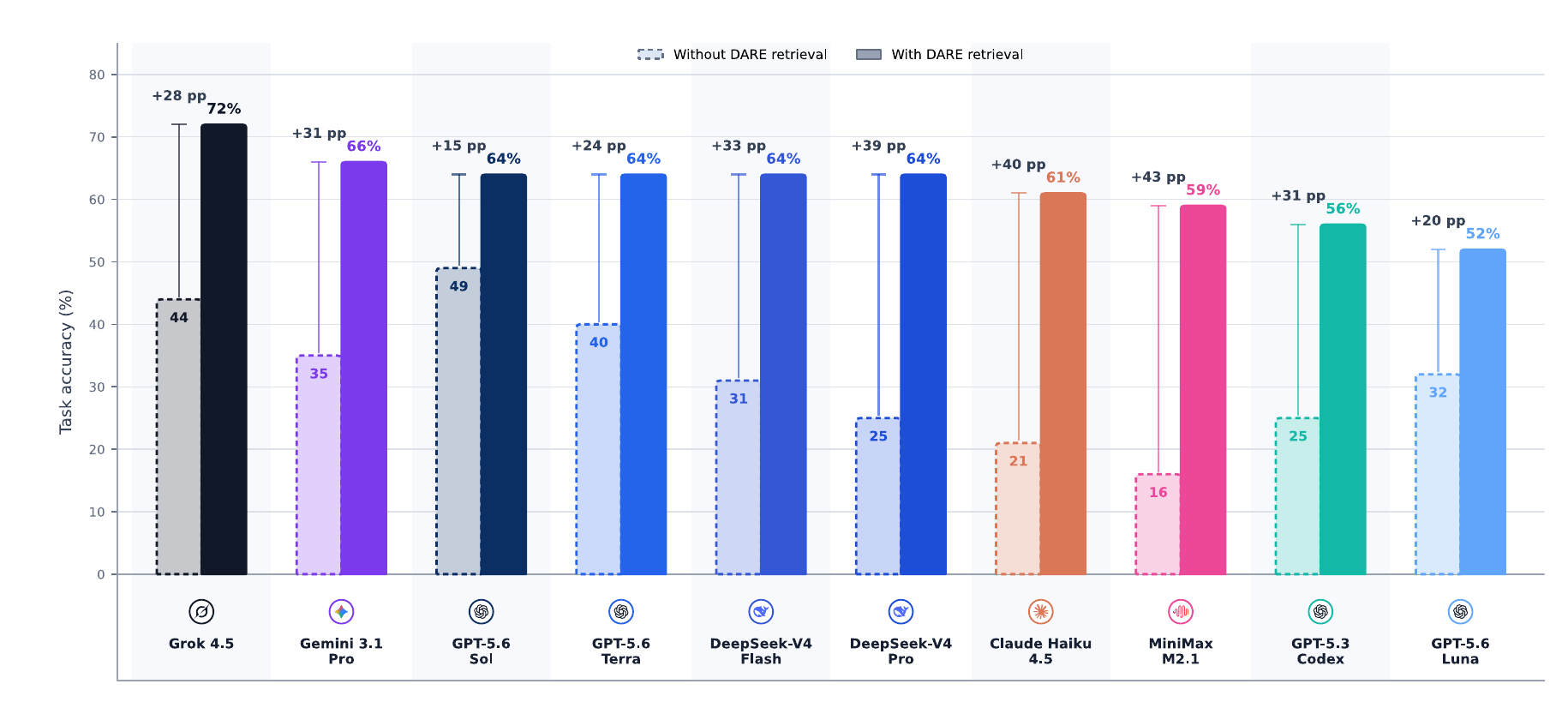}
\caption{Overall accuracy of ten LLMs on RStatBench with and without DARE retrieval.}
\label{fig:overall}
\end{figure*}
We evaluate RCodingAgent with ten LLMs on RStatBench, with and without DARE retrieval. As shown in Figure~\ref{fig:overall}, retrieval improves accuracy for every model and raises the mean from 31.8\% to 62.2\%, a gain of 30.4 percentage points. Across models, accuracy ranges from 16--49\% without retrieval and 52--72\% with retrieval. The largest gains occur for MiniMax-M2.1 (16\% to 59\%), Claude-Haiku-4.5 (21\% to 61\%), and DeepSeek-V4-Pro (25\% to 64\%). Grok-4.5 achieves the highest accuracy with retrieval at 72\%. Thus, DARE benefits models across a wide range of baseline accuracies, although the magnitude of the gain varies.

Figure~\ref{fig:agent_efficiency} compares accuracy and efficiency with and without DARE retrieval; each arrow connects the two conditions for the same model. All ten models achieve higher accuracy and use fewer agent steps with retrieval. Despite the added retrieval step, mean end-to-end duration falls from 107.77 to 65.67 seconds, while mean agent steps decrease from 9.65 to 6.49. Mean token usage also falls from 31,869.62 to 19,700.28 per run (38.18\%), and mean cost per task decreases from \$0.07127 to \$0.04494 (36.94\%). The shorter runs are consistent with agents receiving relevant function guidance early and spending fewer steps exploring alternatives or correcting failed attempts, as discussed further in Section~6. Per-model results and cost details are provided in Supplementary Material Section~L.

\subsection{RQ5: Domain Performance and Failure Analysis}\label{sec:error_analysis}

Figure~\ref{fig:rstatbench_domains} shows task accuracy across statistical domains for each LLM. With DARE retrieval, the radar polygons generally extend farther from the center and enclose substantially larger areas than without retrieval, showing that the accuracy gains span multiple domains rather than a small set of tasks.  The strongest results with DARE occur in Biostatistics and Epidemiology, where mean accuracy rises from 30.77\% to 74.62\%, and Time-series and Signal Analysis, where it rises from 25.56\% to 64.44\%. Performance remains lower in Network and Graphical Models (50.00\%), Econometrics and Social Statistics (55.56\%), and Statistical Genetics and Bioinformatics (58.18\%). The network domain contains only three tasks, so this estimate is less stable; among domains with at least nine tasks, econometrics and statistical genetics present the greatest difficulty. The smaller gain in statistical genetics, from 40.91\% to 58.18\%, also shows that relevant retrieval does not resolve every challenge in specialized analyses.

To see how retrieval changes these outcomes, we analyze agent traces. DARE corrects 333 previously unsuccessful task--model runs. The number that reach the 20-step limit without finishing falls from 162 to 26, while R argument-binding errors fall from 38 to 13. Retrieval can also correct the statistical procedure itself. In Task 4807, DeepSeek-V4-Pro used \texttt{lme4::lmer} and an ad hoc jackknife, producing an incorrect variance-component interval of $[-136.90,435.12]$. With DARE, it used the requested \texttt{minque::lmm.jack} procedure and obtained $[-34.41489,170.38]$, matching the reference after rounding.

Nevertheless, substantial errors remain with DARE retrieval: 378 of the 1,000 runs are incorrect. Most involve the analytical result itself: 209 runs report an incorrect numerical or statistical value, and 113 report an incorrect count, shape, or selection. The latter count is unchanged from the no-retrieval condition. In Task 7694, all ten models fail despite receiving the correct function documentation, reporting RMS or chi-square values from a model configuration different from the one requested. This case exposes a limitation in the agents' statistical reasoning: they have access to the appropriate tool but fail to translate the analytical requirements into the correct model configuration. Supplementary Material Section~M reports failure categories and paired-case analyses.

\section{Discussion and Future Work}

DARE improves R-function retrieval across all nine embedding backbones, while its integration into RCodingAgent raises mean RStatBench accuracy from 31.8\% to 62.2\%. The accompanying reductions in agent steps, token usage, and duration suggest that relevant documentation helps agents choose statistical tools earlier and avoid unsuccessful attempts. Yet tool guidance does not ensure sound statistical reasoning: agents may still use an unsuitable model configuration or report the wrong statistic. Future work could investigate data-aware statistical reasoning to strengthen how LLMs connect observed data characteristics with modeling assumptions, parameter choices, and checks of numerical results.

\begin{strip}
\centering
\includegraphics[width=\textwidth]{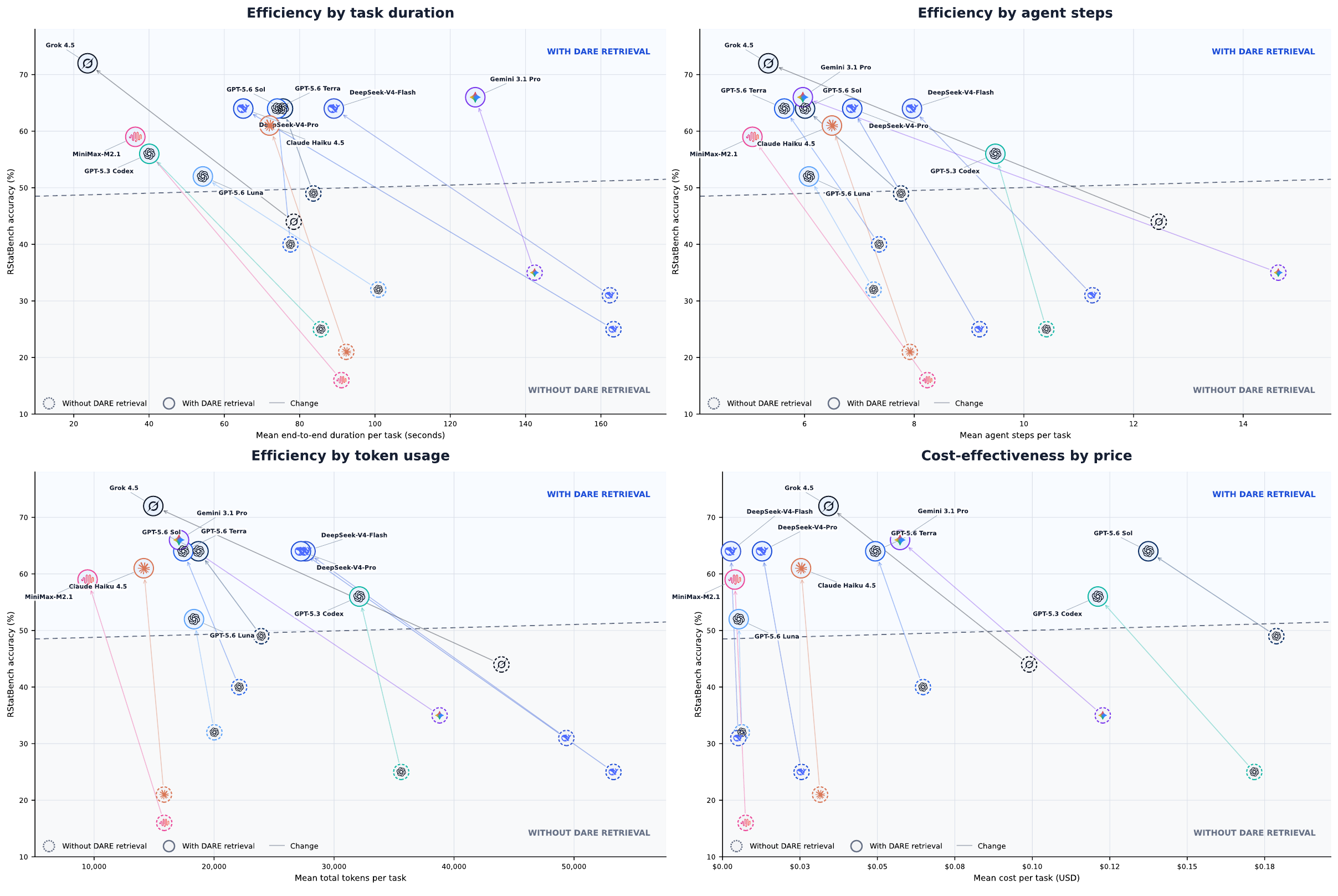}
\captionof{figure}{Efficiency analysis and cost-effectiveness over the agent workflow: (a) end-to-end task duration, (b) agent steps, (c) total token usage, and (d) cost (\$). Each arrow connects the same model without and with DARE retrieval.}
\label{fig:agent_efficiency}
\end{strip}

\begin{strip}
\centering
\includegraphics[width=\textwidth,height=0.78\textheight,keepaspectratio]{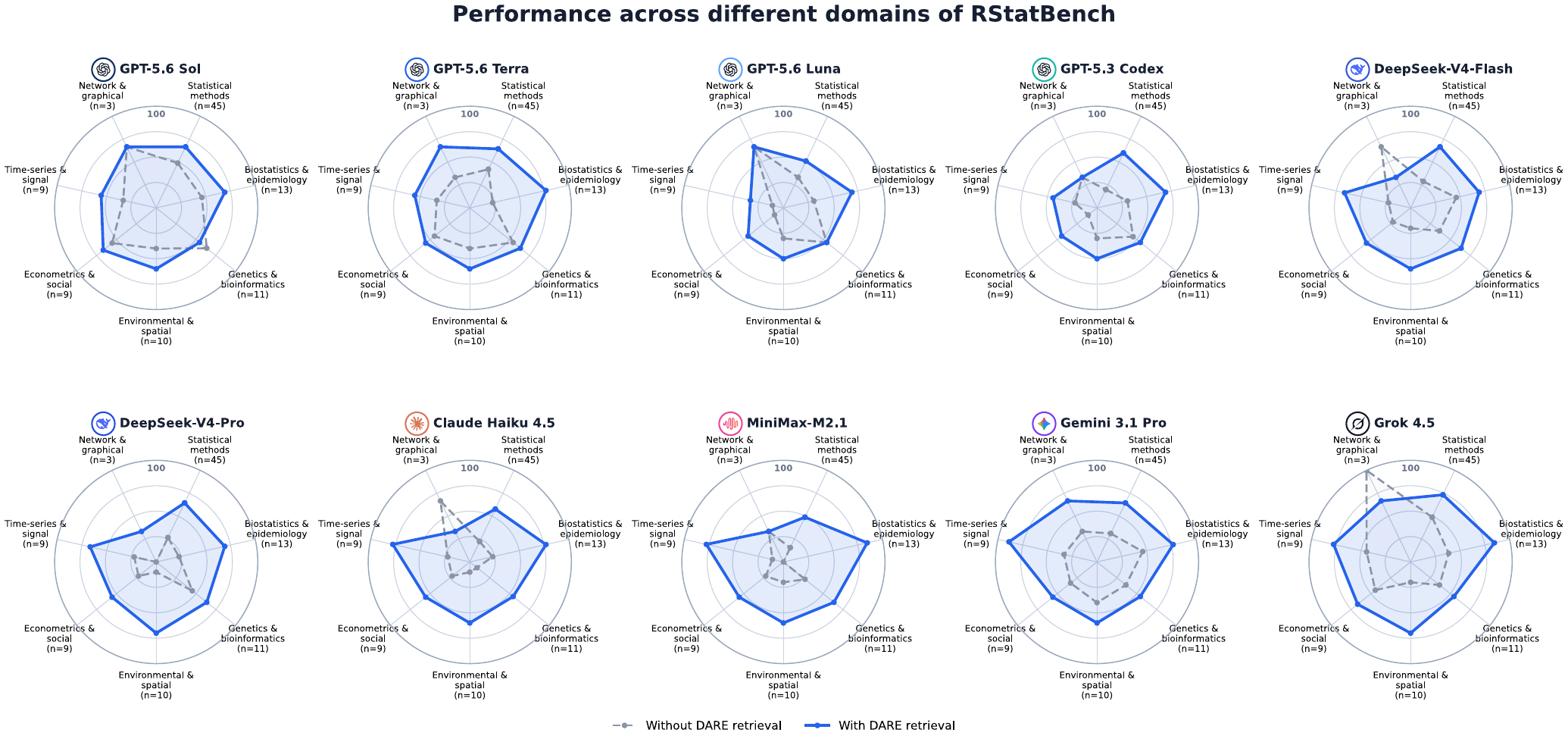}
\captionof{figure}{RStatBench accuracy by application domain for ten LLMs, with and without DARE retrieval. Domain labels give the number of tasks.}
\label{fig:rstatbench_domains}
\end{strip}

The retrieval data also present two limitations. Each query has one source-function label, although other functions may be equally suitable. Collecting multiple relevant functions per query could provide more complete training supervision and support graded-relevance evaluation. In addition, training on a broader range of RPKB functions could improve generalization to previously unseen functions and packages.

Finally, RPKB profiles are inferred from documentation and can contain unsupported attributes. In an audit of 100 functions, 93.52\% of determinate claims in the curated profiles and 90.97--91.86\% in independent regenerations were supported (Supplementary Material Section~K). These errors have not prevented the observed retrieval gains, but may affect individual recommendations. We will maintain RPKB as a dynamic resource through user reports and verified profile corrections.

\section{Conclusion}

We introduce DARE, a distribution-aware representation learning framework that aligns natural-language requests with statistically compatible R functions in RPKB. DARE improves retrieval across all evaluated embedding backbones, while its MiniLM-based instantiation combines high retrieval accuracy with low latency. We further integrate DARE into RCodingAgent and introduce the 100-task RStatBench benchmark. Across ten LLMs, DARE raises mean downstream accuracy from 31.8\% to 62.2\% while reducing agent steps and token usage. These results demonstrate that data-aware retrieval can connect LLM agents more effectively with the established R statistical ecosystem and support more reliable end-to-end analysis.

\bibliographystyle{icml2026}
\bibliography{ref}

@Manual{rcore,
   title = {R: A Language and Environment for Statistical Computing},
   author = {{R Core Team}},
    organization = {R Foundation for Statistical Computing},
   address = {Vienna, Austria},
  year = {2025},
   url = {https://www.R-project.org/},
  }

@article{starcoder,
  title={StarCoder: may the source be with you!},
  author={Li, Raymond and others},
  journal={arXiv:2305.06161},
  year={2023}
}

@inproceedings{karpukhin2020dense,
  title={Dense Passage Retrieval for Open-Domain Question Answering.},
  author={Karpukhin, Vladimir and Oguz, Barlas and Min, Sewon and Lewis, Patrick SH and Wu, Ledell and Edunov, Sergey and Chen, Danqi and Yih, Wen-tau},
  booktitle={EMNLP (1)},
  pages={6769--6781},
  year={2020}
}

@article{izacard2021unsupervised,
  title={Unsupervised dense information retrieval with contrastive learning},
  author={Izacard, Gautier and Caron, Mathilde and Hosseini, Lucas and Riedel, Sebastian and Bojanowski, Piotr and Joulin, Armand and Grave, Edouard},
  journal={arXiv:2112.09118},
  year={2021}
}

@article{toolformer,
  title={Text embeddings by weakly-supervised contrastive pre-training},
  author={Wang, Liang and Yang, Nan and Huang, Xiaolong and Jiao, Binxing and Yang, Linjun and Jiang, Daxin and Majumder, Rangan and Wei, Furu},
  journal={arXiv:2212.03533},
  year={2022}
}

@article{merrick2024embeddingclusteringdataimprove_snowflake_arctic,
      title={Embedding And Clustering Your Data Can Improve Contrastive Pretraining},
      author={Luke Merrick},
      year={2024},
      journal={arXiv:2407.18887},
      url={https://arxiv.org/abs/2407.18887},
}

@inproceedings{intfloat-eg-large,
  title={Improving Text Embedding Models with Positive-aware Hard-negative Mining},
  author={Moreira, Gabriel de Souza P and Osmulski, Radek and Xu, Mengyao and Ak, Ronay and Schifferer, Benedikt and Oldridge, Even},
  booktitle={Proceedings of the 34th ACM International Conference on Information and Knowledge Management},
  pages={2169--2178},
  year={2025}
}

@article{jinaai-jina-embedding,
      title={Jina Embeddings 2: 8192-Token General-Purpose Text Embeddings for Long Documents},
      author={Michael Günther and Jackmin Ong and Isabelle Mohr and Alaeddine Abdessalem and Tanguy Abel and Mohammad Kalim Akram and Susana Guzman and Georgios Mastrapas and Saba Sturua and Bo Wang and Maximilian Werk and Nan Wang and Han Xiao},
      year={2024},
      journal={arXiv:2310.19923},
      url={https://arxiv.org/abs/2310.19923},
}

@inproceedings{bge-m3embedding,
    title = "{M}3-Embedding: Multi-Linguality, Multi-Functionality, Multi-Granularity Text Embeddings Through Self-Knowledge Distillation",
    author = "Chen, Jianlyu  and
      Xiao, Shitao  and
      Zhang, Peitian  and
      Luo, Kun  and
      Lian, Defu  and
      Liu, Zheng",
    booktitle = "Findings of the Association for Computational Linguistics: ACL 2024",
    month = aug,
    year = "2024",
    url = "https://aclanthology.org/2024.findings-acl.137/",
    doi = "10.18653/v1/2024.findings-acl.137",
    pages = "2318--2335"
}

@article{mxbai-large-v1,
  title={Open Source Strikes Bread - New Fluffy Embeddings Model},
  author={Sean Lee and Aamir Shakir and Darius Koenig and Julius Lipp},
  year={2024},
  journal={Availble at: https://www.mixedbread.ai/blog/mxbai-embed-large-v1},
}

@inproceedings{uae-large-v1,
    title = "{A}o{E}: Angle-optimized Embeddings for Semantic Textual Similarity",
    author = "Li, Xianming  and
      Li, Jing",
    booktitle = "Proceedings of the 62nd Annual Meeting of the Association for Computational Linguistics (Volume 1: Long Papers)",
    month = aug,
    year = "2024",
    url = "https://aclanthology.org/2024.acl-long.101/",
    doi = "10.18653/v1/2024.acl-long.101",
    pages = "1825--1839"
}

@inproceedings{gte-large-en-v1.5,
    title = "{mGTE}: Generalized Long-Context Text Representation and Reranking Models for Multilingual Text Retrieval",
    author = "Zhang, Xin  and
      Zhang, Yanzhao  and
      Long, Dingkun  and
      Xie, Wen  and
      Dai, Ziqi  and
      Tang, Jialong  and
      Lin, Huan  and
      Yang, Baosong  and
      Xie, Pengjun  and
      Huang, Fei  and
      Zhang, Meishan  and
      Li, Wenjie  and
      Zhang, Min",
    booktitle = "Proceedings of the 2024 Conference on Empirical Methods in Natural Language Processing: Industry Track",
    month = nov,
    year = "2024",
    publisher = "Association for Computational Linguistics",
    url = "https://aclanthology.org/2024.emnlp-industry.103/",
    doi = "10.18653/v1/2024.emnlp-industry.103",
    pages = "1393--1412"
}

@inproceedings{all-mpnet-base-v2,
author = {Song, Kaitao and Tan, Xu and Qin, Tao and Lu, Jianfeng and Liu, Tie-Yan},
title = {{MPNet}: masked and permuted pre-training for language understanding},
year = {2020},
isbn = {9781713829546},
booktitle = {Proceedings of the 34th International Conference on Neural Information Processing Systems},
pages={16857--16867}
}

@article{sun2026dsaeval,
  title={DSAEval: Evaluating Data Science Agents on a Wide Range of Real-World Data Science Problems},
  author={Sun, Maojun and Xie, Yifei and Wu, Yue and Han, Ruijian and Jiang, Binyan and Sun, Defeng and Yuan, Yancheng and Huang, Jian},
  journal={arXiv:2601.13591},
  year={2026}
}

@article{chen2025largelanguagemodelbaseddata,
      title={Large Language Model-based Data Science Agent: A Survey},
      author={Ke Chen and Peiran Wang and Yaoning Yu and Xianyang Zhan and Haohan Wang},
      journal={arXiv:2508.02744},
      year={2025}
}

@article{zhang2023data,
  title={Data-copilot: Bridging billions of data and humans with autonomous workflow},
  author={Zhang, Wenqi and Shen, Yongliang and Lu, Weiming and Zhuang, Yueting},
  journal={arXiv:2306.07209},
  year={2023}
}

@article{zhang2025deepanalyze,
  title={Deepanalyze: Agentic large language models for autonomous data science},
  author={Zhang, Shaolei and Fan, Ju and Fan, Meihao and Li, Guoliang and Du, Xiaoyong},
  journal={arXiv:2510.16872},
  year={2025}
}

@inproceedings{hong2025data,
  title={Data interpreter: An llm agent for data science},
  author={Hong, Sirui and Lin, Yizhang and Liu, Bang and Liu, Bangbang and Wu, Binhao and Zhang, Ceyao and Li, Danyang and Chen, Jiaqi and Zhang, Jiayi and Wang, Jinlin and others},
  booktitle={Findings of the Association for Computational Linguistics: ACL 2025},
  pages={19796--19821},
  year={2025}
}

@article{nelder1972generalized,
  title={Generalized linear models},
  author={Nelder, John Ashworth and Wedderburn, Robert WM},
  journal={Journal of the Royal Statistical Society Series A: Statistics in Society},
  volume={135},
  number={3},
  pages={370--384},
  year={1972},
  publisher={Oxford University Press}
}

@article{dezeure2015high,
  title={High-dimensional inference: confidence intervals, p-values and R-software hdi},
  author={Dezeure, Ruben and B{\"u}hlmann, Peter and Meier, Lukas and Meinshausen, Nicolai},
  journal={Statistical Science},
  pages={533--558},
  year={2015},
  publisher={JSTOR}
}

@article{rai2021survival,
  title={Survival analysis: A primer for the clinician scientists},
  author={Rai, Sushmita and Mishra, Prabhakar and Ghoshal, Uday C},
  journal={Indian Journal of Gastroenterology},
  volume={40},
  number={5},
  pages={541--549},
  year={2021},
  publisher={Springer}
}

@article{Lin03042021,
author = {Kevin Z. Lin and Jing Lei and Kathryn Roeder},
title = {Exponential-Family Embedding With Application to Cell Developmental Trajectories for Single-Cell RNA-Seq Data},
journal = {Journal of the American Statistical Association},
volume = {116},
number = {534},
pages = {457--470},
year = {2021},
publisher = {Taylor \& Francis},
doi = {10.1080/01621459.2021.1886106},
URL = {
        https://doi.org/10.1080/01621459.2021.1886106
},
eprint = {
        https://doi.org/10.1080/01621459.2021.1886106
}
}

@article{Burman01122006,
author = {Prabir Burman and Robert H Shumway},
title = {Generalized Exponential Predictors for Time Series Forecasting},
journal = {Journal of the American Statistical Association},
volume = {101},
number = {476},
pages = {1598--1606},
year = {2006},
publisher = {Taylor \& Francis},
doi = {10.1198/016214506000000483},
URL = {https://doi.org/10.1198/016214506000000483},
eprint = {https://doi.org/10.1198/016214506000000483}
}

@article{ihaka1996r,
  title={R: a language for data analysis and graphics},
  author={Ihaka, Ross and Gentleman, Robert},
  journal={Journal of computational and graphical statistics},
  volume={5},
  number={3},
  pages={299--314},
  year={1996},
  publisher={Taylor \& Francis}
}

@article{lewis2020retrieval,
  title={Retrieval-augmented generation for knowledge-intensive nlp tasks},
  author={Lewis, Patrick and Perez, Ethan and Piktus, Aleksandra and Petroni, Fabio and Karpukhin, Vladimir and Goyal, Naman and K{\"u}ttler, Heinrich and Lewis, Mike and Yih, Wen-tau and Rockt{\"a}schel, Tim and others},
  journal={Proceedings of the 34th International Conference on Neural Information Processing Systems},
  pages={9459--9474},
  year={2020}
}

@inproceedings{orlanski2023measuring,
  title={Measuring the impact of programming language distribution},
  author={Orlanski, Gabriel and Xiao, Kefan and Garcia, Xavier and Hui, Jeffrey and Howland, Joshua and Malmaud, Jonathan and Austin, Jacob and Singh, Rishabh and Catasta, Michele},
  booktitle={Proceedings of the 40th International Conference on Machine Learning},
  pages={26619--26645},
  year={2023},
  organization={PMLR}
}

@article{shen2023hugginggpt,
  title={Hugginggpt: Solving ai tasks with chatgpt and its friends in hugging face},
  author={Shen, Yongliang and Song, Kaitao and Tan, Xu and Li, Dongsheng and Lu, Weiming and Zhuang, Yueting},
  journal={Proceedings of the 37th Advances in Neural Information Processing Systems},
  pages={38154--38180},
  year={2023}
}

@article{reimers2019sentence,
  title={Sentence-bert: Sentence embeddings using siamese bert-networks},
  author={Reimers, Nils and Gurevych, Iryna},
  journal={arXiv:1908.10084},
  year={2019}
}

@article{oord2019representationlearningcontrastivepredictive,
      title={Representation Learning with Contrastive Predictive Coding},
      author={Aaron van den Oord and Yazhe Li and Oriol Vinyals},
      journal={arXiv:1807.03748},
      year={2019},
      url={https://arxiv.org/abs/1807.03748}
}

@article{zhao2025current,
  title={Do current language models support code intelligence for r programming language?},
  author={Zhao, Zixiao and Fard, Fatemeh},
  journal={ACM Transactions on Software Engineering and Methodology},
  volume={34},
  number={8},
  pages={1--39},
  year={2025},
  publisher={ACM New York, NY}
}

@inproceedings{adamw,
  author       = {Ilya Loshchilov and
                  Frank Hutter},
  title        = {Decoupled Weight Decay Regularization},
  booktitle    = {7th International Conference on Learning Representations, {ICLR} 2019,
                  New Orleans, LA, USA, May 6-9, 2019},
  publisher    = {OpenReview.net},
  year         = {2019},
  url          = {https://openreview.net/forum?id=Bkg6RiCqY7}
}

@article{sun2026rejoinder1,
  title={Rejoinder to the Discussions on “LAMBDA: A Large Model Based Data Agent”},
  author={Sun, Maojun and Han, Ruijian and Jiang, Binyan and Qi, Houduo and Sun, Defeng and Yuan, Yancheng and Huang, Jian},
  journal={Journal of the American Statistical Association},
  volume={121},
  number={553},
  pages={36--43},
  year={2026},
  publisher={Taylor \& Francis}
}

@article{sun2026rejoinder2,
  title={Rejoinder to the Discussions on “A Survey on Large Language Model-based Agents for Statistics and Data Science” A Statistical View of LLM-Based Data Science Agents: Reasoning, Uncertainty, and Reliability},
  author={Sun, Maojun and Han, Ruijian and Jiang, Binyan and Qi, Houduo and Sun, Defeng and Yuan, Yancheng and Huang, Jian},
  journal={The American Statistician},
  volume={80},
  number={3},
  pages={352--359},
  year={2026},
  publisher={Taylor \& Francis}
}

@article{sun2026survey,
  title={A survey on large language model-based agents for statistics and data science},
  author={Sun, Maojun and Han, Ruijian and Jiang, Binyan and Qi, Houduo and Sun, Defeng and Yuan, Yancheng and Huang, Jian},
  journal={The American Statistician},
  volume={80},
  number={3},
  pages={331--344},
  year={2026},
  publisher={Taylor \& Francis}
}

@article{sun2026lambda,
  title={Lambda: A large model based data agent},
  author={Sun, Maojun and Han, Ruijian and Jiang, Binyan and Qi, Houduo and Sun, Defeng and Yuan, Yancheng and Huang, Jian},
  journal={Journal of the American Statistical Association},
  volume={121},
  number={553},
  pages={1--13},
  year={2026},
  publisher={Taylor \& Francis}
}

\clearpage
\onecolumn
\appendix
\section*{Supplementary Material}
This supplement presents the data-profile and query-generation prompts and an example RPKB record (Sections A--C), followed by experimental settings and training trajectories (Section D). It then describes the RCodingAgent prompts and agent loop (Section E), RStatBench construction and representative tasks, and a paired agent execution example (Sections F--I). Sections J--M report retrieval results under alternative training and test splits, profile and relevance audits, agent efficiency and cost, and downstream failure analysis.

\section{Data-Profile Generation Prompt}
\label{supp:dpg}
This section presents the prompt used to convert R documentation into structured Data Profiles. It instructs an
LLM to exclude generic utilities and extract key statistical metadata such as data modality, distribution assumptions, and
dimensionality.

\begin{promptbox}{Example Prompt of Data Profile Generation}
You are an expert Data Scientist and R Developer. Your task is to filter and annotate R functions for a Retrieval-Augmented Generation (DARE) system specialized in data science. \\

You will receive a JSON describing an R function (including package description, function description, usage, arguments).\\

Your goal is to:\\
1. FILTER: Determine if this function is a specific algorithm, statistical method, data transformation, or machine learning model. \\
   - REJECT basic utilities (e.g., file I/O, simple string manipulation, getters/setters, pure plotting without analytics).\\
   - KEEP only if it has clear data input requirements and analytical purpose.\\

2. EXTRACT: If the function is relevant, extract structured Data Information. Specifically, what kind of data does this algorithm expect? (e.g., sparse matrices, categorical factors, normally distributed vectors).\\

Output must be a valid JSON object adhering to this schema:\\
\{\\
  "is\_relevant": boolean, \\
  "data\_profile": \{\\
    "data\_modality": "string [tabular, time-series, text, image, graph, genomic/sequence, other]",\\
    "feature\_type": "string [numerical, categorical, mixed, binary, text-token, any]",\\
    "distribution\_assumption": "string (e.g., 'normal', 'sparse', 'poisson', 'any', 'unknown')",\\
    "dimensionality": "string [low, high, any]",\\
    "missing\_data\_handling": "string (e.g., 'handles\_na', 'must\_be\_complete', 'unknown')",\\
    "specific\_constraints": ["list", "of", "strings"]\\
  \},\\
  "task\_type": "string (e.g., 'classification', 'clustering', 'normalization', 'gene\_signature\_scoring')",\\
  "reasoning": "short explanation"\\
\}\\

\end{promptbox}

\section{DARE Training Query-Generation Prompt}
\label{supp:tqg}
The prompt below generates 30 queries per function in five styles: detailed, problem-focused, help-me, instruction-first, and short keyword requests. 

\begin{promptbox}{DARE Training Query-Generation Prompt}
You are an expert Data Scientist creating a training dataset for a Semantic Search Engine.\\
Your goal is to generate **30 diverse user search queries** that would lead a user to the target R function described below.\\

**CRITICAL RULES:**\\
1. **NEVER mention the function name or the package name** in the queries. The user doesn't know the tool exists yet; they are describing their problem.\\
2. **HEAVILY RELY on the Data Information**: The queries MUST describe the data structure, distribution, or constraints provided in the context (e.g., "I have a sparse matrix...", "My data is high-dimensional...", "Handling zero-inflated count data...").\\
3. **Variety**:\\
   - 12 queries should be long and detailed (e.g., "I am working with a large tabular dataset that has high sparsity, and I need to...").\\
   - 5 queries should be problem-focused (e.g., "How to handle missing values in categorical time-series?").\\
   - 5 queries should be "Help me" style (e.g., "I need a way to plot gene expressions where the data is...").\\
   - 3 queries should be Instruction first then followed by data information (e.g. "I need to visualize the structure of protein... Data Information: {'type': 'tabular', 'distribution': 'sparse'.....} ")\\
   - 5 queries should be short/keyword-based but include data types (e.g., "dim reduction for sparse arrays").\\

**Input Format:** Function Description + Data Profile JSON.\\
**Output Format:** A valid JSON object with a single key "queries" containing a list of 30 strings.\\
Example: {"queries": ["query 1", "query 2", ...]}\\

\end{promptbox}

\section{Example of RPKB}
\label{supp:erpkb}
The following examples illustrate RPKB records containing function documentation and data profiles.

\begin{promptbox}{Example of RPKB}
\begin{lstlisting}
{
  "id": 22,
  "fc_id": "Reproducible_Data_Screening_Checks_and_Report_of
  _Possible_Errors::makeDataReport",
  "ground_truth_doc": "Data Constraints: {\"data_modality\": \"tabular\", \"feature_type\": \"mixed\", \"distribution_assumption\": \"any\", \"dimensionality\": \"any\", \"missing_data_handling\": \"handles_na\", \"specific_constraints\": [\"data.frame/tibble /matrix\", \"handles labelled/haven_labelled as factors\", \"smartNum treats low unique nums as factors\"]}. Task: data_quality_assessment. \nR Package: Reproducible_Data_Screening_Checks_and_Report_of_Possible Errors. R Package Description: Data screening is an important first step of any statistical\n    analysis. 'dataReporter' auto generates a customizable data report with a thorough\n    summary of the checks ......
  "data_profile": {
      "data_modality": "tabular",
      "feature_type": "mixed",
      "distribution_assumption": "any",
      "dimensionality": "any",
      "missing_data_handling": "handles_na",
      "specific_constraints": [
          "data.frame/tibble/matrix",
          "handles labelled/haven_labelled as factors",
          "smartNum treats low unique nums as factors"
      ]
  },
  "usage_guidance": "makeCodebook(data, vol = \"\", reportTitle = NULL, file = NULL, ...)",
  "example_code":    "library(dataReporter)
                      data(testData)
                      data(toyData)
                      check(toyData)
                      DF <- data.frame(x = 1:15)
                      makeDataReport(DF)
                      data(testData)
                      makeDataReport(testData)
                      # Overwrite any existing files generated by makeDataReport
                      makeDataReport(testData, replace=TRUE)
                      # Change output format to Word/docx:
                      makeDataReport(testData, replace=TRUE, output = "word")
                      # Only include problematic variables in the output document
                      makeDataReport(testData, replace=TRUE, onlyProblematic=TRUE)
                      # Add user defined check-function to the checks performed on character variables:
                      # Here we add functionality to search for the string wally (ignoring case)

                      wheresWally <- function(v, ...) {
                      res <- grepl("wally", v, ignore.case=TRUE)
                      problem <- any(res)
                      message <- "Wally was found in these data"
                      checkResult(list(problem = problem,
                      message = message,
                      problemValues = v[res]))
                      }
                      wheresWally <- checkFunction(wheresWally,
                      description = "Search for the string 'wally' ignoring case",
                      classes = c("character")
                      )
                      # Add the newly defined function to the list of checks used for characters.
                      makeDataReport(testData,
                      checks = setChecks(character = defaultCharacterChecks(add = "wheresWally")),
                      replace=TRUE)
                      #Handle non-supported variable classes using treatXasY: treat raw as character and
                      #treat complex as numeric. We also add a list variable, but as lists are not
                      #handled through treatXasY, this variable will be caught in the preChecks and skipped:
                      toyData$rawVar <- as.raw(c(1:14, 1))
                      toyData$compVar <- c(1:14, 1) + 2i
                      toyData$listVar <- as.list(c(1:14, 1))
                      makeDataReport(toyData, replace = TRUE,
                      treatXasY = list(raw = "character", complex = "numeric"))",
  "task_type": "data_quality_assessment",
  "domain": "Specialized Statistics & ML"
}
\end{lstlisting}
\end{promptbox}

\begin{promptbox}{Example of RPKB}
\begin{lstlisting}
{
  "id": 960,
  "fc_id": "OPTICS_K-Xi_Density-Based_Clustering::fortify_ica",
  "ground_truth_doc": "Data Constraints: {\"data_modality\": \"tabular\", \"feature_type\":  \"numerical\", \"distribution_assumption\": \"non-gaussian\", \"dimensionality\": \"high\", \"missing_data_handling\": \"must_be_complete\", \"specific_constraints\": [\"numeric matrix\", \"suitable for blind source separation\"]}. Task: independent_component_analysis. \nR Package: OPTICS_K-Xi_Density-Based_Clustering. R Package Description: Density-based clustering methods are well adapted to the clustering of high-dimensional data and enable the discovery of core groups of various shapes despite large amounts of noise. This package provides a novel density-based cluster extraction method, OPTICS k-Xi, and a framework to compare k-Xi models using ......
      "feature_type": "numerical",
      "distribution_assumption": "non-gaussian",
      "dimensionality": "high",
      "missing_data_handling": "must_be_complete",
      "specific_constraints": [
          "numeric matrix",
          "suitable for blind source separation"
      ]
  },
  "usage_guidance": "fortify_ica(m_data, ..., sup_vars = NULL)",
  "example_code": "library(opticskxi)
                      library(fastICA)
                      library(plyr)
                      df_ica <- fortify_ica(iris[-5], n.comp = 2)
                      df_ica",
  "task_type": "independent_component_analysis",
  "domain": "Specialized Statistics & ML"
}
\end{lstlisting}
\end{promptbox}

\begin{promptbox}{Example of RPKB}
\begin{lstlisting}
{
    "id": 2460,
    "fc_id": "Estimate_a_Log-Concave_Probability_Density_from_I
    id_Observations::logconTwoSample",
    "ground_truth_doc": "Data Constraints: {\"data_modality\": \"tabular\", \"feature_type\": \"numerical\", \"distribution_assumption\": \"log-concave\", \"dimensionality\": \"low\", \"missing_data_handling\": \"must_be_complete\", \"specific_constraints\": [\"univariate iid observations\", \"two samples for comparison\"]}. Task: two-sample hypothesis test.  ......
    "data_profile": {
        "data_modality": "tabular",
        "feature_type": "numerical",
        "distribution_assumption": "log-concave",
        "dimensionality": "low",
        "missing_data_handling": "must_be_complete",
        "specific_constraints": [
            "univariate iid observations",
            "two samples for comparison"
        ]
    },
    "usage_guidance": "logconTwoSample(x, y, which = c("MLE", "smooth"), M = 999,
                        n.grid = 500, display = TRUE, seed0 = 1977)
                        ",
    "example_code": "library(logcondens)
                    n1 <- 30
                    n2 <- 25
                    x <- rgamma(n1, 2, 1)
                    y <- rgamma(n2, 2, 1) + 1
                    twosample <- logconTwoSample(x, y, which = c("MLE", "smooth")[1], M = 999)
                    ",
    "task_type": "two-sample hypothesis test",
    "domain": "Specialized Statistics & ML"
}
\end{lstlisting}
\end{promptbox}

\section{Training and Evaluation Details}
\label{supp:loss}
We use MiMo-V2-flash to generate 30 queries per source function in five styles: twelve detailed, five problem-focused, five help-me, three instruction-first, and five short keyword queries. The finalized dataset contains 245,738 queries, of which 212,974 are assigned to training and 32,764 to testing under the query-level split. Each query has one source-function label, and retrieval is evaluated against the same corpus of 8,191 functions.

\begin{figure}[tbp]
\centering
\includegraphics[width=\textwidth]{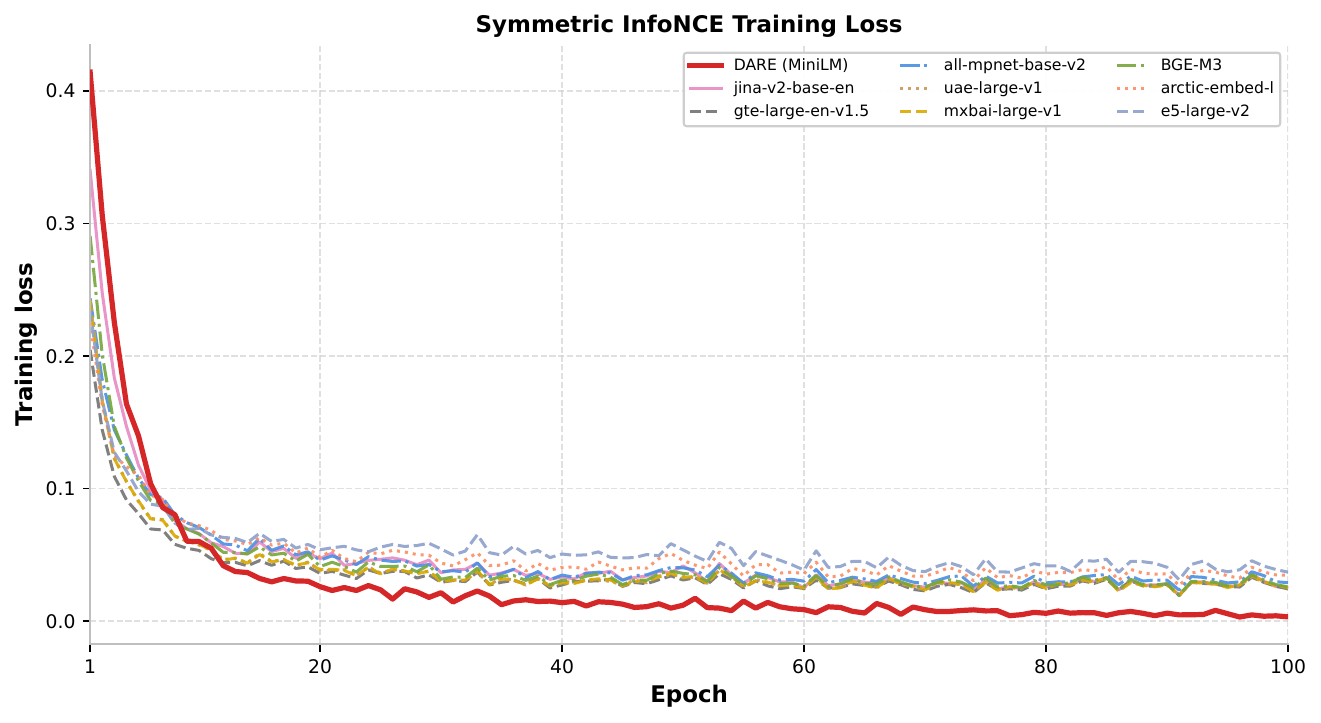}
\caption{Training loss for the 100-epoch runs.}
\label{supp:training_loss}
\end{figure}

\begin{figure}[tbp]
\centering
\includegraphics[width=\textwidth]{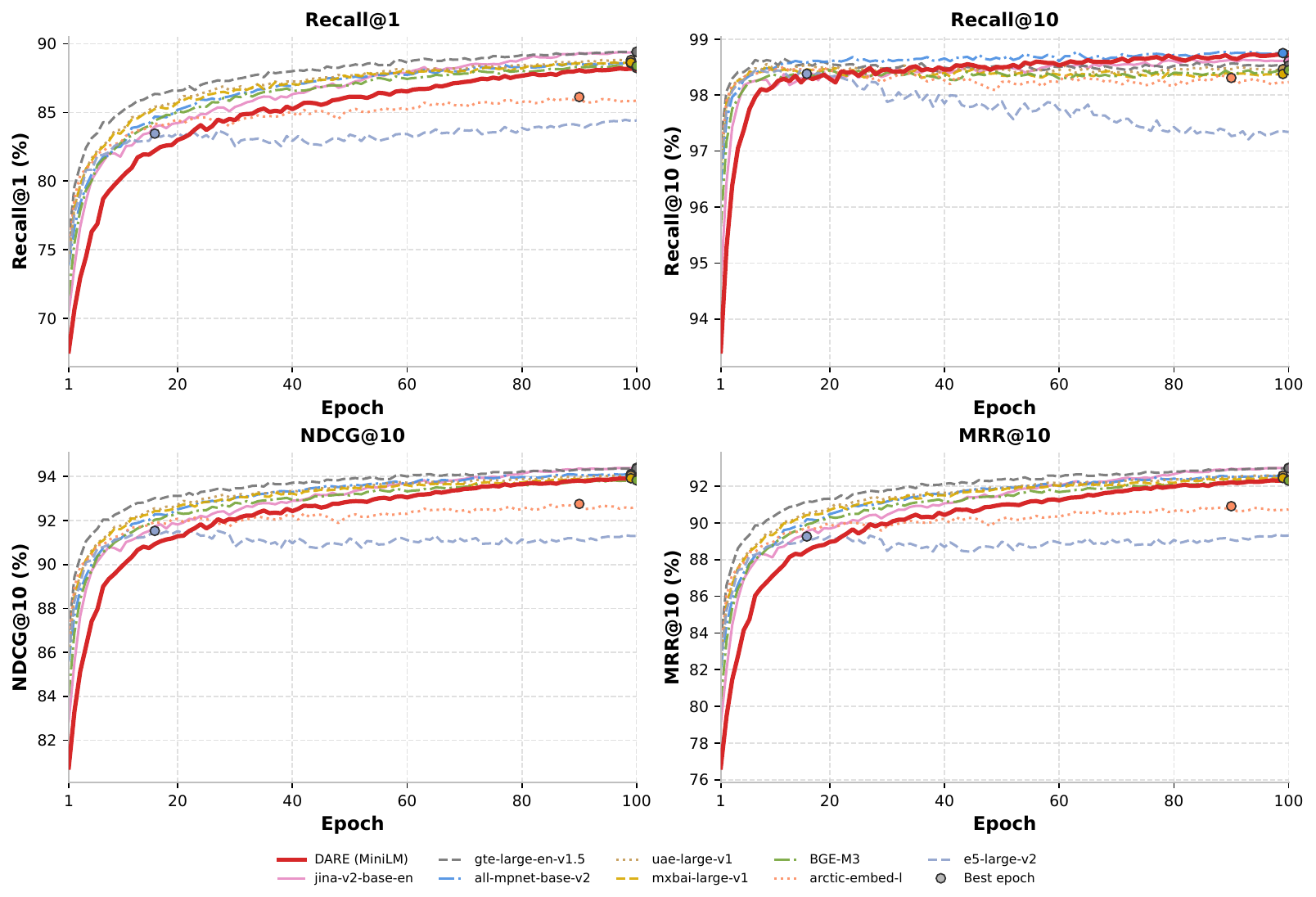}
\caption{Retrieval metrics at each checkpoint for the nine fine-tuned backbones.}
\label{supp:curves}
\end{figure}

All nine backbones are trained for 100 epochs with a batch size of 256, AdamW, BF16, and symmetric InfoNCE on the same RPKB query-level split. MiniLM and Jina use a learning rate of $10^{-4}$ with 10\% warmup. Applying these settings to the other seven backbones led to unstable training or representation collapse; their final runs therefore use a learning rate of $2\times10^{-5}$, 1\% warmup, and gradient-norm clipping at 1.0. The log inverse temperature is initialized to 2.659, with the inverse temperature capped at 100. Model-specific input prefixes are used where required. Figure~\ref{supp:training_loss} shows the loss and Figure~\ref{supp:curves} demonstrates recall@1, recall@10, NDCG@10, MRR@10 during the training process.

\section{RCodingAgent Prompts and Agent Loop}
\label{supp:spra}
RCodingAgent uses the same R-execution workflow with and without retrieval. The two system prompts below specify the agent instructions for each condition; only the DARE-enabled condition provides retrieved R documentation before code generation.

\begin{promptbox}{Example System Prompt For RCodingAgent}
You are an advanced Data Science Agent specializing in R. \\
Your goal is to solve complex data analysis tasks by writing robust, efficient, and "tidy" R code.

\textbf{Core Guidelines:} \\
\textbf{1. File I/O Safety:} \\
- \textbf{Input:} specific data files are located at \texttt{\{data\_path\}}. ONLY load data from this directory. \\
- \textbf{Output:} save all results (CSVs, models, images) to \texttt{\{working\_path\}}. \\
\textbf{2. Package Management:} Install any packages if you need. \\
\textbf{3. Task Execution:} Solving the problem by multiple step, but finally you only need to give a concise answer with numerical results. \\
\textbf{4. System Integrity:} \textbf{IMPORTANTLY:} You cannot perform any actions that may harm the system, such as deleting directories, important software, or software packages.
\end{promptbox}

\begin{promptbox}{Example System Prompt For RCodingAgent with DARE}
You are an advanced Data Science Agent specializing in R equipped with a R package retrieval system.

\textbf{Your Ability:} \\
You will be provided with \textbf{"Retrieved R Documentation"}. Use these documents to solve the task. \\
- You MUST carefully analyze the "Retrieved R Documentation". \\
- If the documents provide a specific function, \textbf{prioritize using it} over general knowledge. \\
- Mimic the usage style, syntax, and pay attention to "example" or "usage" sections.

\textbf{Core Guidelines:} \\
\textbf{1. File I/O Safety:} Input from \texttt{\{data\_path\}}, output to \texttt{\{working\_path\}}. \\
\textbf{2. Task Execution:} Solve the problem by multiple steps, and give a concise answer with numerical results. \\
\textbf{3. System Integrity:} Do not perform actions that harm the system.
\end{promptbox}

For DARE retrieval, an optional query rewrite combines the user's request with information extracted from the uploaded dataset before functions are ranked. The rewriting prompt is shown below.

\begin{promptbox}{Query-Rewriting Prompt}
Rewrite the following user query using the corresponding data information. Preserve the user's analytical intent and make data characteristics relevant to selecting an appropriate R function explicit. If the query conflicts with the data information, follow the observed data. Do not invent unsupported data properties. Return only the rewritten query.\\[0.5em]
\textbf{Query:} \texttt{\{Query\}}\\
\textbf{Data Information:} \texttt{\{Data profile\}}\\
\textbf{Rewritten Query:} 
\end{promptbox}

\subsection{RCodingAgent Framework and Agent Loop}
\label{supp:agent_loop}
\begin{figure}[!htbp]
  \centering
  \includegraphics[width=\linewidth]{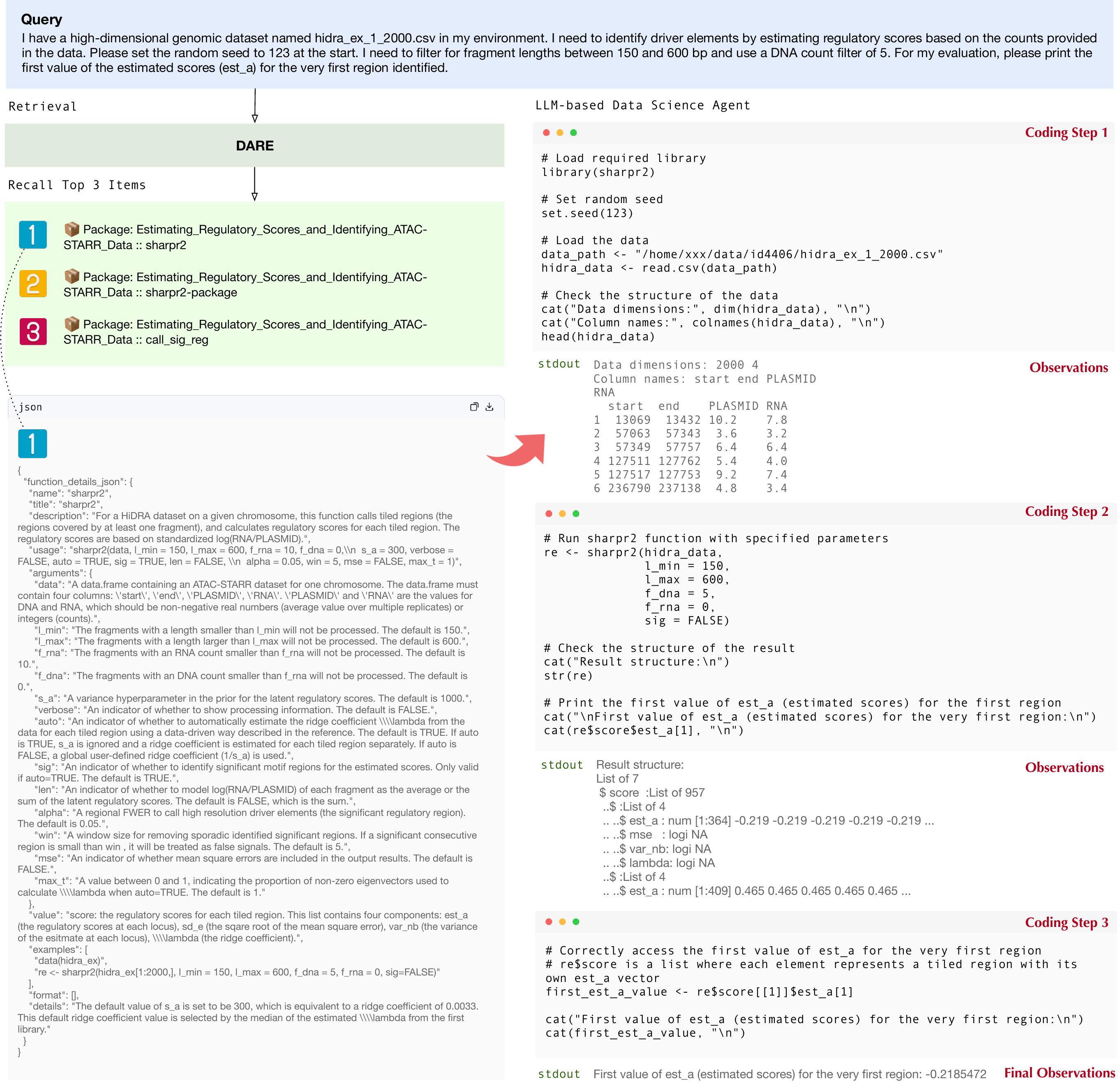}
  \caption{RCodingAgent execution for a HiDRA regulatory-score task, showing retrieved functions, R code, and observations across agent steps.}
  \label{fig:supp_dare_agent}
\end{figure}

Algorithm~\ref{alg:agent} specifies how retrieval and R execution are integrated. When retrieval is enabled, monitored data information may be incorporated into the request, and the top three RPKB function records are added to the agent context before its first action. The agent then alternates R code generation with sandbox execution, using each observation to guide its next step. Figure~\ref{fig:supp_dare_agent} illustrates this workflow for the regulatory-score task examined in Section~\ref{supp:ec}.

\begin{algorithm}[!htbp]
\begin{singlespace}
\caption{Agent Loop for RCodingAgent. Here, $\widetilde q$ is the retrieval query, $H_t$ is the interaction history, $a_t$ is the LLM response at step $t$, $C_t$ is the extracted R code, $o_t$ is the sandbox observation, and $R$ is the final report.}
\label{alg:agent}
\small
\begin{algorithmic}[1]
\Require Request $q$, optional dataset $c$, switches $b,w\in\{0,1\}$
\Require Corpus $\mathcal{F}$, encoder $\varepsilon$, $K=3$, step limit $T=20$
\Ensure Final report and artifacts, or a non-completion record
\State $H_0\gets\operatorname{Initialize}(q,c)$
\State $\widetilde q\gets q$
\If{$b=1\land c\neq\varnothing$}
    \State $c_q\gets\operatorname{Monitor}(c)$
    \If{$w=1$}
        \State $\widetilde q\gets\operatorname{Rewrite}_{\mathrm{LLM}}(q,c_q)$
    \Else
        \State $\widetilde q\gets[q;c_q]$
    \EndIf
\EndIf
\If{$b=1$}
    \State $\mathcal{K}\gets\operatorname{TopK}_{f=(d,c_d)\in\mathcal{F}}
    s(\varepsilon(\widetilde q),\varepsilon([d;c_d]))$
    \State $H_0\gets\operatorname{Append}(H_0,\operatorname{Context}(\mathcal K))$
\EndIf
\For{$t=1,\ldots,T$}
    \State $a_t\gets\operatorname{LLM}(H_{t-1})$
    \If{$a_t$ contains a final report}
        \State \Return final report and generated artifacts
    \EndIf
    \State $C_t\gets\operatorname{ExtractRCode}(a_t)$
    \State $o_t\gets\operatorname{Execute}_{R}(C_t,c)$
    \State $H_t\gets\operatorname{Append}(H_{t-1},a_t,C_t,o_t)$
\EndFor
\State \Return $\operatorname{NonCompletion}(H_T)$
\end{algorithmic}
\end{singlespace}
\end{algorithm}

\section{Construction of RStatBench}
\label{supp:aet}
RStatBench tasks are grounded in executable R analyses with associated data and verifiable numerical outputs. The prompt below converts each task's statistical objective and function metadata into user-facing requests while withholding the package and function names.

\begin{promptbox}{RStatBench Task-Query Generation Prompt}
\textbf{Role:} You are an expert in R programming and advanced statistics assisting in a research evaluation. Your task is to generate natural language queries based on R function metadata to test the retrieval and execution capabilities of an AI Agent. \\

\textbf{Core Objective:} \\
The queries must never mention the R package or function names. Instead, they must deeply embed the Distribution Assumptions, Data Modalities, and Specific Constraints found in the metadata. The goal is to force the model to retrieve the correct tool based on "statistical logic" rather than "name memorization." \\

\textbf{Task Instructions (4 Query Styles):} \\
\textbf{1. Beginner Style (Vague):} Focus on the high-level scientific goal without technical jargon. \\
\textbf{2. Practitioner Style (Rough):} Use standard industry terminology and describe the operational context. \\
\textbf{3. Rigorous Researcher Style (Detailed):} Explicitly include statistical profiles (e.g., Poisson, Gaussian, Time-series). This is the core test for DARE's performance. \\
\textbf{4. Technical Logic Style (Parametric):} Emphasize specific parameter configurations and the exact output required for evaluation. \\

\textbf{Hard Constraints for Output:} \\
\textbf{- Mandatory Seed:} Every query must require the user to start the script with \texttt{set.seed(123)}. \\
\textbf{- Mandatory Verification:} Every query must instruct the agent to print a specific value (the Ground Truth) for evaluation (e.g., "print the p-value" or "output the first value of \texttt{est\_a}"). \\
\textbf{- Pre-loaded Data:} Assume the dataset is already available. The Agent must read the data directly (e.g., from a CSV or a specific object name). \\
\textbf{- No Leakage:} Strictly forbid mentioning the \texttt{fc\_id} or the R Package name in the query. \\

\textbf{Example Format:} \\
"I have a dataset [Data Modality] that follows a [Distribution Assumption]. My goal is to [Task Type]. Please write an R script starting with \texttt{set.seed(123)} to process the data... Finally, print [Ground Truth Metric] so I can verify the results."
\end{promptbox}

\section{Example RStatBench Task}
\label{supp:esat}
Task 960 illustrates how one statistical objective can be expressed at different levels of specificity. The example records the data profile and reference output for independent component analysis, followed by four requests that retain the same target computation.
\begin{promptbox}{ Example (ID: 960)}
\textbf{Task Type:} Independent Component Analysis \\
\textbf{Data Profile:} Tabular, Numerical, Non-Gaussian Distribution, High-dimensional. \\

\textbf{Ground Truth Verification:} \\
\texttt{df\_ica\$IC1[1] = 0.5040647} \\
\texttt{df\_ica\$IC2[1] = -1.372758} \\

\textbf{Generated Evaluation Queries:} \\
\textbf{1. Beginner Style:} I have a file named \texttt{iris\_data.csv} containing multiple numeric measurements. I suspect there are several independent underlying signals mixed together in this data, and I want to separate them into 2 distinct components. Can you write an R script that extracts these signals and adds them as new columns to the original data? Please set the seed to 123 for consistency and show me the first row of the two newly created columns. \\

\textbf{2. Practitioner Style:} I need to perform a feature extraction on \texttt{iris\_data.csv} to find independent components. Please use R to reduce the numeric features to 2 components. I'd like the results to be integrated back into the dataset (fortified). Make sure to \texttt{set.seed(123)} at the start and print the first row of the components so I can verify the values. \\

\textbf{3. Rigorous Researcher Style:} I am working with \texttt{iris\_data.csv}, which is a high-dimensional numeric matrix. Given that the data follows a non-Gaussian distribution, I need to perform Independent Component Analysis (ICA) for blind source separation. Please extract 2 components and ensure they are fortified into the data frame. Set the random seed to 123 and output the first row of ICA1 and ICA2 to confirm the computation. \\

\textbf{4. Technical Logic Style:} Using the numeric data in \texttt{iris\_data.csv}, I need to run a density-based ICA extraction to get 2 components. It's crucial that the method supports a 'fortify' mechanism to merge the components back into the observation table automatically. Use seed 123 for reproducibility and display the top row of the resulting ICA columns.
\end{promptbox}

\section{Representative RStatBench Task Sources}
\label{supp:16aet}
RStatBench includes analyses that require specialized functions from several R packages. Table~\ref{tab:task_profiles} identifies 16 representative tasks, the functions used in their reference solutions, and the corresponding package documentation.

\begin{singlespace}
\begin{small}
    \renewcommand{\arraystretch}{1.5}
    \setlength{\tabcolsep}{8pt}
    \begin{longtable}{p{0.28\linewidth} p{0.64\linewidth}}
        \caption{Representative RStatBench tasks and functions used in their reference solutions.\label{tab:task_profiles}} \\
        \hline
        \textbf{Statistical Task} & \textbf{Detailed Information} \\
        \hline
        \endfirsthead

        \hline
        \textbf{Statistical Task} & \textbf{Detailed Information} \\
        \hline
        \endhead

        \hline
        \endfoot

        \textbf{Independent component analysis} & \textbf{Package/Function:} OPTICS K-Xi Density-Based Clustering::fortify\_ica \\
        & \textbf{Description:} Density-based clustering utilities and visualization for ICA results \\
        & \textbf{Link:}  \url{https://cran.r-project.org/web/packages/opticskxi/}
		vignettes/opticskxi.pdf \\
        \hline
        \textbf{Two-sample hypothesis test} & \textbf{Package/Function:} Estimate a Log-Concave Probability Density from Iid Observations::logconTwoSample \\
        & \textbf{Description:} Two-sample hypothesis testing based on log-concave density estimation. \\
        & \textbf{Link:} \url{https://cran.r-project.org/web/packages/logcondens/logcondens.pdf} \\
        \hline
        \textbf{Goodness-of-fit} & \textbf{Package/Function:} National Information Platforms for Nutrition Anthropometric Data Toolkit::ageChildren \\
        & \textbf{Description:} Anthropometric age processing and child age standardization utilities. \\
        & \textbf{Link:} \url{https://cran.r-project.org/web/packages/nipnTK/nipnTK.pdf} \\
        \hline
        \textbf{Age heaping detection} & \textbf{Package/Function:} National Information Platforms for Nutrition Anthropometric Data Toolkit::ageHeaping \\
        & \textbf{Description:} Detection and correction of age heaping phenomena in demographic data \\
        & \textbf{Link:} \url{https://cran.r-project.org/web/packages/nipnTK/nipnTK.pdf} \\
        \hline
        \textbf{Narrow endemic identification} & \textbf{Package/Function:} Auxiliary Functions to Estimate Centers of Biodiversity::getNarrowEndemics \\
        & \textbf{Description:} Identification of narrow endemic species for biodiversity analysis. \\
        & \textbf{Link:} \url{https://cran.r-project.org/web/packages/sperich/sperich.pdf} \\
        \hline
        \textbf{Environmental affinity assignment} & \textbf{Package/Function:} Diversity Dynamics using Fossil Sampling Data::affinity \\
        & \textbf{Description:} Environmental affinity assignment for fossil occurrence records \\
        & \textbf{Link:} \url{https://cran.r-project.org/web/packages/divDyn/divDyn.pdf} \\
        \hline
        \textbf{Survivorship analysis} & \textbf{Package/Function:} Diversity Dynamics using Fossil Sampling Data::survivors \\
        & \textbf{Description:} Survivorship analysis across temporal fossil sampling intervals \\
        & \textbf{Link:} \url{https://cran.r-project.org/web/packages/divDyn/divDyn.pdf} \\
        \hline
        \textbf{Conditional graphical model estimation} & \textbf{Package/Function:} Conditional Graphical LASSO (cglasso) \\
        & \textbf{Description:} Estimation of conditional graphical models under censoring and missingness \\
        & \textbf{Link:} \url{https://cran.r-project.org/web/packages/cglasso/cglasso.pdf} \\
        \hline
        \textbf{Clinical EHR preprocessing} & \textbf{Package/Function:} Electronic Health Record Data Processing and Analysis Tool::processLastDose \\
        & \textbf{Description:} Temporal standardization and preprocessing of clinical dosage records. \\
        & \textbf{Link:} \url{https://cran.r-project.org/web/packages/EHR/EHR.pdf} \\
        \hline
        \textbf{Clinical trial simulation} & \textbf{Package/Function:} Clinical Trial Simulation::cut\_data\_by\_event \\
        & \textbf{Description:} Survival data truncation based on event occurrence time. \\
        & \textbf{Link:} \url{https://cran.r-project.org/web/packages/simtrial/simtrial.pdf} \\
        \hline
        \textbf{Regulatory score estimation} & \textbf{Package/Function:} Estimating Regulatory Scores and Identifying ATAC-STARR Data::sharpr2 \\
        & \textbf{Description:} Regulatory activity score estimation from high-throughput genomic assays \\
        & \textbf{Link:} \url{https://cran.r-project.org/web/packages/sharpr2/sharpr2.pdf} \\
        \hline
        \textbf{Ancestry score computation} & \textbf{Package/Function:} Computation of Ancestry Scores with Mixed Families and Unrelated Individuals::perfectwhiten \\
        & \textbf{Description:} Covariance-preserving whitening for ancestry score computation. \\
        & \textbf{Link:} \url{https://cran.r-project.org/web/packages/PCFAM/PCFAM.pdf}\\
        \hline
        \textbf{Linear mixed model} & \textbf{Package/Function:} Various Linear Mixed Model Analyses::lmm.jack \\
        & \textbf{Description:} Linear mixed-effects modeling with jackknife-based inference \\
        & \textbf{Link:} \url{https://cran.r-project.org/web/packages/minque/minque.pdf} \\
        \hline
        \textbf{Gaussian mixture overlap} & \textbf{Package/Function:} Simulating Data to Study Performance of Clustering Algorithms::overlap \\
        & \textbf{Description:} Computation of Gaussian mixture overlap for clustering evaluation. \\
        & \textbf{Link:} \url{https://cran.r-project.org/web/packages/MixSim/MixSim.pdf} \\
        \hline
        \textbf{Factor Analysis Preprocessing} & \textbf{Package/Function:} Set of Assumptions for Factor and Principal Component Analysis::communalities \\
        & \textbf{Description:} Preprocessing and communality estimation for factor analysis. \\
        & \textbf{Link:} \url{https://cran.r-project.org/web/packages/FactorAssumptions/FactorAssumptions.pdf} \\
        \hline
        \textbf{Financial quality scoring} & \textbf{Package/Function:} Quality Scores for the Russell 3000::market\_profitability \\
        & \textbf{Description:} Profitability z-scoring and financial quality metric computation. \\
        & \textbf{Link:} \url{https://cran.r-project.org/web/packages/qmj/qmj.pdf} \\
        \hline
    \end{longtable}
\end{small}
\end{singlespace}

\section{Paired RCodingAgent Execution Example}
\label{supp:ec}
Task 4406 asks the agent to estimate regulatory scores from a HiDRA dataset and report the first score for the first region. The reference value is $-0.2185$. We compare executions with and without DARE retrieval to examine how access to function documentation affects the statistical procedure and extraction of the requested result.

\begin{promptbox}{Query and Ground Truth}
\textbf{User Query:} I have a high-dimensional genomic dataset named hidra\_ex\_1\_2000.csv in my environment. I need to identify driver elements by estimating regulatory scores based on the counts provided in the data. Please set the random seed to 123 at the start. I need to filter for fragment lengths between 150 and 600 bp and use a DNA count filter of 5. For my evaluation, please print the first value of the estimated scores (est\_a) for the very first region identified.

\textbf{Ground Truth:} -0.2185
\end{promptbox}

\subsection{Retrieved function}
DARE ranks \texttt{sharpr2} first for this request. Its RPKB record describes the required HiDRA data structure, filtering arguments, and regulatory-score output, providing the agent with the specialized procedure before code generation.
\begin{promptbox}{DARE-Retrieved Top Three Functions}
\begin{lstlisting}[basicstyle=\ttfamily\tiny,breaklines=true]
{
 1. [ID: Estimating_Regulatory_Scores_and_Identifying_ATAC-STARR_Data::sharpr2] (Dist: 0.7704)
    Package: Estimating_Regulatory_Scores_and_Identifying_ATAC-STARR_Data :: sharpr2
    {'function_name': 'sharpr2', 'package_version': '1.1.1.0', 'package_name': 'Estimating_Regulatory_Scores_and_Identifying_ATAC-STARR_Data', 'task_type': 'regulatory_score_estimation', 'data_profile_json': '{"data_modality": "genomic/sequence", "feature_type": "numerical", "distribution_assumption": "poisson", "dimensionality": "high", "missing_data_handling": "must_be_complete", "specific_constraints": ["ATAC-STARR dataset", "columns: start, end, PLASMID, RNA", "non-negative reals or integers", "one chromosome", "fragment lengths 150-600 bp", "RNA/DNA count filters"]}', 'function_details_json': '{"name": "sharpr2", "title": "sharpr2", "description": "For a HiDRA dataset on a given chromosome, this function calls tiled regions (the regions covered by at least one fragment), and calculates regulatory scores for each tiled region. The regulatory scores are based on standardized log(RNA/PLASMID).", "usage": "sharpr2(data, l_min = 150, l_max = 600, f_rna = 10, f_dna = 0,\\n  s_a = 300, verbose = FALSE, auto = TRUE, sig = TRUE, len = FALSE, \\n  alpha = 0.05, win = 5, mse = FALSE, max_t = 1)", "arguments": {"data": "A data.frame containing an ATAC-STARR dataset for one chromosome. The data.frame must contain four columns: \'start\', \'end\', \'PLASMID\', \'RNA\'. \'PLASMID\' and \'RNA\' are the values for DNA and RNA, which should be non-negative real numbers (average value over multiple replicates) or integers (counts).", "l_min": "The fragments with a length smaller than l_min will not be processed. The default is 150.", "l_max": "The fragments with a length larger than l_max will not be processed. The default is 600.", "f_rna": "The fragments with an RNA count smaller than f_rna will not be processed. The default is 10.", "f_dna": "The fragments with an DNA count smaller than f_rna will not be processed. The default is 0.", "s_a": "A variance hyperparameter in the prior for the latent regulatory scores. The default is 1000.", "verbose": "An indicator of whether to show processing information. The default is FALSE.", "auto": "An indicator of whether to automatically estimate the ridge coefficient \\\\lambda from the data for each tiled region using a data-driven way described in the reference. The default is TRUE. If auto is TRUE, s_a is ignored and a ridge coefficient is estimated for each tiled region separately. If auto is FALSE, a global user-defined ridge coefficient (1/s_a) is used.", "sig": "An indicator of whether to identify significant motif regions for the estimated scores. Only valid if auto=TRUE. The default is TRUE.", "len": "An indicator of whether to model log(RNA/PLASMID) of each fragment as the average or the sum of the latent regulatory scores. The default is FALSE, which is the sum.", "alpha": "A regional FWER to call high resolution driver elements (the significant regulatory region). The default is 0.05.", "win": "A window size for removing sporadic identified significant regions. If a significant consecutive region is small than win , it will be treated as false signals. The default is 5.", "mse": "An indicator of whether mean square errors are included in the output results. The default is FALSE.", "max_t": "A value between 0 and 1, indicating the proportion of non-zero eigenvectors used to calculate \\\\lambda when auto=TRUE. The default is 1."}, "value": "score: the regulatory scores for each tiled region. This list contains four components: est_a (the regulatory scores at each locus), sd_e (the sqare root of the mean square error), var_nb (the variance of the esitmate at each locus), \\\\lambda (the ridge coefficient).", "examples": ["data(hidra_ex)", "re <- sharpr2(hidra_ex[1:2000,], l_min = 150, l_max = 600, f_dna = 5, f_rna = 0, sig=FALSE)"], "format": [], "details": "The default value of s_a is set to be 300, which is equivalent to a ridge coefficient of 0.0033. This default ridge coefficient value is selected by the median of the estimated \\\\lambda from the first library."}', 'function_title': 'sharpr2'}

   2. [ID: Estimating_Regulatory_Scores_and_Identifying_ATAC-STARR_Data::sharpr2-package] (Dist: 1.0781)
      Package: Estimating_Regulatory_Scores_and_Identifying_ATAC-STARR_Data :: sharpr2-package
{'function_name': 'sharpr2-package', 'task_type': 'regulatory score estimation and driver element identification', 'data_profile_json': '{"data_modality": "genomic/sequence", "feature_type": "mixed", "distribution_assumption": "sparse", "dimensionality": "high", "missing_data_handling": "unknown", "specific_constraints": ["ATAC-STARR sequencing data", "high-definition reporter assay library datasets"]}', 'function_details_json': '{"name": "sharpr2-package", "title": "Estimating regularoty scores and identifying high resolution driver elements for ATAC-STARR data", "description": "The package develops an algorithm for identifying high-resolution driver elements for datasets from an ATAC-STARR library.", "usage": "", "arguments": {}, "value": "", "examples": ["data(hidra_ex)", "re <- sharpr2(hidra_ex[1:2000,], l_min = 150, l_max = 600, f_dna = 5, f_rna = 0, sig=FALSE)"], "format": [], "details": ""}', 'package_version': '1.1.1.0', 'function_title': 'Estimating regularoty scores and identifying high resolution driver elements for ATAC-STARR data', 'package_name': 'Estimating_Regulatory_Scores_and_Identifying_ATAC-STARR_Data'}


   3. [ID: Estimating_Regulatory_Scores_and_Identifying_ATAC-STARR_Data::call_sig_reg] (Dist: 1.1114)
      Package: Estimating_Regulatory_Scores_and_Identifying_ATAC-STARR_Data :: call_sig_reg
{'function_title': 'call_sig_reg', 'data_profile_json': '{"data_modality": "genomic/sequence", "feature_type": "numerical", "distribution_assumption": "normal", "dimensionality": "high", "missing_data_handling": "unknown", "specific_constraints": ["requires sharpr2 output object", "tiled genomic regions", "z-score thresholds"]}', 'package_name': 'Estimating_Regulatory_Scores_and_Identifying_ATAC-STARR_Data', 'function_name': 'call_sig_reg', 'task_type': 'significant_regulatory_region_calling', 'package_version': '1.1.1.0', 'function_details_json': '{"name": "call_sig_reg", "title": "call_sig_reg", "description": "Given an object returned from the sharpr2 function, this function calls significant regions that contain driver elements for a specific tiled region based on a user-defined threshold.", "usage": "call_sig_reg(res, nr, threshold = 3.5, win = 10)", "arguments": {"res": "An object obtained from the sharpr2 function.", "nr": "An integer indicating the number of tiled region in res for which driver elements will be called.", "threshold": "The cutoff to identify driver elements in the tiled region. The positions with a z-score larger than the threshold will be called. The default is 3.5.", "win": "A window size for removing sporadic significant regions. If a significant consecutive region is small than win, it will be treated as false signals. The default is 10."}, "value": "sig_reg: identified regions containing driver elements.", "examples": ["data(hidra_ex)", "re  <- sharpr2(hidra_ex[1:2000,], l_min = 150, l_max = 600, f_dna = 5, f_rna = 0, sig=TRUE)", "call_sig_reg(re,850, threshold=2.5)"], "format": [], "details": ""}'}

}
\end{lstlisting}
\end{promptbox}

\subsection{Execution with DARE retrieval}
The retrieval-enabled agent applies \texttt{sharpr2} with the requested fragment-length and DNA-count filters. After inspecting the nested output structure, it reports $-0.2185472$, which agrees with the reference value after rounding. The execution trace below shows how the agent selects the method and extracts the result.

%
%
%
%
%
%
%

\begin{promptbox}{Code Log: Task ID 4406 }
	\ttfamily\small 
	
	\textbf{[Agent] step 0: call LLM} \\
	\textbf{[Agent] tool\_call} $\to$ \texttt{jupyter\_code\_interpreter}
	\begin{lstlisting}[style=agentcode]
		# Set random seed and load specialized library
		set.seed(123)
		library(sharpr2)
		
		# Load the genomic dataset
		data_path <- "/home/xxx/id4406/hidra_ex_1_2000.csv"
		hidra_data <- read.csv(data_path)
		
		# Display basic info
		cat("Dataset dimensions:", dim(hidra_data), "\n")
		head(hidra_data)
	\end{lstlisting}
	\textit{[Tool Output] Dataset dimensions: 2000 4 \\
		\begin{tabular}{llll}
			start & end & PLASMID & RNA \\
			13069 & 13432 & 10.2 & 7.8 \\
			... & ... & ... & ...
	\end{tabular}}
	
	\vspace{0.8em} \hrule \vspace{0.8em} 
	
	\textbf{[Agent] step 1: call LLM} \\
	\textbf{[Agent] tool\_call} $\to$ \texttt{jupyter\_code\_interpreter}
	\begin{lstlisting}[style=agentcode]
		# Run sharpr2 with distribution-specific parameters
		result <- sharpr2(
		data = hidra_data, l_min = 150, l_max = 600,
		f_rna = 0, f_dna = 5, sig = FALSE, verbose = TRUE
		)
		
		cat("\nResult names:", names(result), "\n")
		cat("Length of est_a:", length(result$score$est_a), "\n")
	\end{lstlisting}
	\textit{[Tool Output] Total reads after filtering: 1326. Call tiled regions: 957. Calculating regulatory scores ... Done! \\
		Result names: score region n\_reg n\_read sig\_reg motif cutoff \\
		Length of est\_a: 0}
	
	\vspace{0.8em} \hrule \vspace{0.8em}
	
	\textbf{[Agent] step 2: call LLM} \\
	\textbf{[Agent] tool\_call} $\to$ \texttt{jupyter\_code\_interpreter}
	\begin{lstlisting}[style=agentcode]
		# Examine nested structure for correct value extraction
		cat("$score structure:\n")
		str(result$score[[1]])
	\end{lstlisting}
	\textit{[Tool Output] \$score structure: \\
		List of 4 \\
		.. \$ est\_a : num [1:364] -0.219 -0.219 -0.219 -0.219 -0.219 ... \\
		.. \$ mse   : logi NA}
	
	\vspace{0.8em} \hrule \vspace{0.8em}
	
	\textbf{[Agent] step 3: call LLM} \\
	\textbf{[Agent] tool\_call} $\to$ \texttt{jupyter\_code\_interpreter}
	\begin{lstlisting}[style=agentcode]
		# Extract the first value of est_a for the first region
		first_value <- result$score[[1]]$est_a[1]
		print(first_value)
	\end{lstlisting}
	\textit{[Tool Output] [1] -0.2185472}
	
	\vspace{0.8em} \hrule \vspace{0.8em}
	
	\textbf{[Agent] step 4: call LLM} \\
	\textbf{[Agent] final answer:} \\
	Identified 957 tiled regions using \texttt{sharpr2}. The first estimated regulatory score (\texttt{est\_a}) for the first region is \textbf{-0.2185472}.
\end{promptbox}

\subsection{Execution without DARE retrieval}
Without the retrieved documentation, the agent instead calculates a generic log-ratio from the filtered counts. This code executes, but its reported value of $-0.3479233$ does not match the reference score because it does not apply the requested regulatory-score estimator.

\begin{promptbox}{Code Log: Task ID 4406 (Failed Case - Without DARE)}
	\ttfamily\small 
	
	\textbf{[Agent] step 0: call LLM} \\
	\textbf{[Agent] tool\_call} $\to$ \texttt{jupyter\_code\_interpreter}
	\begin{lstlisting}[style=agentcode]
		# Load required libraries and set seed
		set.seed(123)
		
		# Load the data and explore its structure
		data <- read.csv("/home/xxx/id4406/hidra_ex_1_2000.csv")
		dim(data)
		head(data)
	\end{lstlisting}
	\textit{[Tool Output] [1] 2000 4 \\
		\begin{tabular}{llll}
			start & end & PLASMID & RNA \\
			13069 & 13432 & 10.2 & 7.8 \\
			... & ... & ... & ...
	\end{tabular}}
	
	\vspace{0.8em} \hrule \vspace{0.8em} 
	
	\textbf{[Agent] step 1: call LLM} \\
	\textbf{[Agent] tool\_call} $\to$ \texttt{jupyter\_code\_interpreter}
	\begin{lstlisting}[style=agentcode]
		# Calculate fragment length and apply filters manually
		data$fragment_length <- data$end - data$start
		
		# Filter: lengths between 150-600 and PLASMID >= 5
		data_filtered <- data[data$fragment_length >= 150 &
		data$fragment_length <= 600 &
		data$PLASMID >= 5, ]
		dim(data_filtered)
	\end{lstlisting}
	\textit{[Tool Output] [1] 1428 5}
	
	\vspace{0.8em} \hrule \vspace{0.8em}
	
	\textbf{[Agent] step 2: call LLM} \\
	\textbf{[Agent] tool\_call} $\to$ \texttt{jupyter\_code\_interpreter}
	\begin{lstlisting}[style=agentcode]
		# Estimate scores using a generic log2 ratio heuristic
		# Lacking specialized tool documentation for sharpr2
		data_filtered$est_a <- log2((data_filtered$RNA + 1) /
		(data_filtered$PLASMID + 1))
		
		# Print the first value for the first region
		cat("First estimated score (est_a):", data_filtered$est_a[1], "\n")
	\end{lstlisting}
	\textit{[Tool Output] First estimated score (est\_a): -0.3479233}
	
	\vspace{0.8em} \hrule \vspace{0.8em}
	
	\textbf{[Agent] step 3: call LLM} \\
	\textbf{[Agent] final answer:} \\
	I have estimated the regulatory scores by applying filters and calculating a log2 ratio. The first value of the estimated score (\texttt{est\_a}) for the first region is \textbf{-0.3479233}.
\end{promptbox}

\section{Additional Data Splits Results}
\label{supp:split_details}
To examine retrieval beyond the query-level split, we evaluate DARE when functions, packages, or query styles are absent from training. Table~\ref{tab:supp_splits} compares these settings with the 32,764-query query-level test. The function-level split holds out 1,229 functions and evaluates 37,025 queries; the package-level split holds out 111 packages and evaluates 35,278 queries. Both use the full 8,191-function retrieval corpus, so the test functions remain candidates even though their associated queries are absent from training.

For each held-out-style experiment, queries in one generated style are excluded from training and reserved for evaluation. The long-detailed and instruction-first styles yield 80,700 and 20,175 test queries, respectively; the problem-focused, help-me, and short-keyword styles each yield 33,625. 

\begin{table}[htbp]
\begin{singlespace}
\centering\small
\caption{Retrieval performance under query-, function-, package-, and style-level splits. All metrics are percentages.}
\label{tab:supp_splits}
\resizebox{\textwidth}{!}{\begin{tabular}{lrrrr}
\toprule
Setting & NDCG@10 & MRR@10 & Recall@10 & Recall@1\\
\midrule
Query-level split & 93.92 & 92.33 & 98.71 & 88.23\\
\midrule
Function-level split & 85.08 & 81.52 & 95.90 & 72.74\\
Package-level split & 85.75 & 82.71 & 95.06 & 75.48\\
\midrule
Held-out style: long detailed & 93.07 & 91.23 & 98.64 & 86.51\\
Held-out style: problem focused & 92.34 & 90.51 & 97.87 & 85.94\\
Held-out style: help me & 94.73 & 93.39 & 98.79 & 89.91\\
Held-out style: instruction first & 93.55 & 92.07 & 98.09 & 88.42\\
Held-out style: short keyword & 92.41 & 90.82 & 97.28 & 86.97\\
\bottomrule
\end{tabular}}
\end{singlespace}
\end{table}

\section{Profile and Retrieval Audits}
\label{supp:audit_details}
We assess two potential sources of uncertainty in the retrieval evaluation: the accuracy of generated function profiles and the relevance of the source-function labels. We also describe the statistical-compatibility tests and the evaluation using human-authored queries.
\subsection{Documentation support for function profiles}
Because RPKB profiles are generated from package documentation, we assess whether their claims are supported by that documentation. We randomly sample 100 functions and compare their curated profiles with versions independently generated by GPT-5.6-Luna, GPT-5.6-Terra, and GPT-5.6-Sol under the same protocol. A statistical Ph.D. student evaluates claims in all four profiles against the complete function documentation, classifying each as supported, contradicted, or indeterminate. Table~\ref{tab:supp_profile_audit} reports the claim counts and support rates for each source.

\begin{table}[htbp]
\begin{singlespace}
\centering\small
\caption{Human audit of documentation support across the four profile sources.}
\label{tab:supp_profile_audit}
\resizebox{\textwidth}{!}{%
\begin{tabular}{lrrrrrr}
\toprule
Profile source & Claims & Supported & Contradicted & Indeterminate & \shortstack{Support rate\\(all claims, \%)} & \shortstack{Support rate\\(determinate, \%)} \\
\midrule
Original RPKB profile & 889 & 808 & 56 & 25 & 90.89 & 93.52 \\
GPT-5.6-Luna & 960 & 839 & 77 & 44 & 87.40 & 91.59 \\
GPT-5.6-Terra & 910 & 786 & 78 & 46 & 86.37 & 90.97 \\
GPT-5.6-Sol & 916 & 801 & 71 & 44 & 87.45 & 91.86 \\
\bottomrule
\end{tabular}}
\end{singlespace}
\end{table}

\subsection{Compatibility-stress construction}
To distinguish sensitivity to statistical requirements from reliance on shared terminology, we construct two test-query subsets: one varies the distribution within an analytical task, and the other varies the task among functions with similar data profiles. A separate contradiction test pairs misleading distributional wording with correct data information matched on task, modality, and feature type. All tests use the same fine-tuned checkpoint and 8,191-function corpus. For reference, pairing the noncontradictory target query with the same correct data information yields 98.68\% NDCG@10 and 96.43\% Recall@1.

\subsection{Source-label relevance}
To assess the relevance of source functions to synthetic queries, we retrieve ten candidates for each of 100 queries and present their documentation to a validator in randomized order, without scores or source labels. The validator selects three functions per query. The source function appears among those selections for 93 queries, corresponding to a human-audited Recall@3 of 93.0\%.

\subsection{Human-authored queries}
To assess retrieval beyond synthetic queries, we evaluate statistical-analysis requests authored by a statistician during prior interactions with ChatGPT. Of 50 collected requests, 45 are retained after screening for a suitable function in RPKB. A retrieved function is considered relevant if it supports the requested analysis without an explicit incompatibility with the query. Table~\ref{tab:supp_human_queries} reports retrieval performance on the retained queries. 

\begin{table}[!htbp]
\begin{singlespace}
\centering\small
\caption{Retrieval performance on 45 human-authored statistical-analysis queries.}
\label{tab:supp_human_queries}
\begin{tabular}{lr}
\toprule
Metric & Result \\
\midrule
Recall@1 & 84.44\% (38/45) \\
Recall@3 & 97.78\% (44/45) \\
Recall@10 & 100.00\% (45/45) \\
\bottomrule
\end{tabular}
\end{singlespace}
\end{table}

\section{RCodingAgent Efficiency and Cost Analysis}
\label{supp:efficiency}
To assess whether retrieval changes the resources required to complete an analysis, we evaluate RCodingAgent with ten LLMs on all 100 RStatBench tasks, with and without DARE. Both conditions use the same agent configuration and R environment, yielding 1,000 runs per condition. Table~\ref{tab:agent_efficiency} reports end-to-end duration, including model-provider latency and R execution, and the number of agent steps for each model.


\begin{table}[!htbp]
\begin{singlespace}
\centering
\small
\caption{Mean end-to-end duration and agent steps for each model, with and without DARE retrieval.}
\label{tab:agent_efficiency}
\resizebox{0.85\textwidth}{!}{%
\begin{tabular}{lrrrr}
\toprule
& \multicolumn{2}{c}{Mean duration (s)} & \multicolumn{2}{c}{Mean agent steps} \\
\cmidrule(lr){2-3}\cmidrule(lr){4-5}
Model & Without DARE & With DARE & Without DARE & With DARE \\
\midrule
gpt-5.6-sol & 83.64 & 75.58 & 7.76 & 6.01 \\
gpt-5.6-terra & 77.55 & 74.02 & 7.36 & 5.63 \\
gpt-5.6-luna & 100.88 & 54.29 & 7.26 & 6.08 \\
gpt-5.3-codex & 85.63 & 40.06 & 10.41 & 9.48 \\
deepseek-v4-flash & 162.39 & 89.08 & 11.25 & 7.96 \\
deepseek-v4-pro & 163.35 & 65.02 & 9.19 & 6.87 \\
claude-haiku-4-5 & 92.38 & 72.00 & 7.92 & 6.50 \\
minimax-m2.1 & 91.08 & 36.36 & 8.24 & 5.05 \\
gemini-3.1-pro & 142.41 & 126.64 & 14.64 & 5.97 \\
grok-4.5 & 78.43 & 23.67 & 12.46 & 5.34 \\
\midrule
Mean & 107.77 & \textbf{65.67} & 9.65 & \textbf{6.49} \\
\bottomrule
\end{tabular}%
}
\end{singlespace}
\end{table}
To compare token cost across models, we apply the input and output rates in Table~\ref{supp:prices} to each run's recorded token counts. The calculation excludes cached-input discounts and adjustments for long contexts or peak/off-peak pricing. 

\begin{table}[!htbp]
\begin{singlespace}
\centering\small
\caption{Prices used for the cost analysis, in USD per million tokens.}
\label{supp:prices}
\begin{tabular}{lrr}
\toprule
Model & Input & Output\\
\midrule
Claude-Haiku-4.5 & 1.00 & 5.00\\
DeepSeek-V4-Flash & 0.09 & 0.18\\
DeepSeek-V4-Pro & 0.435 & 0.87\\
Gemini-3.1-Pro & 2.00 & 12.00\\
GPT-5.3-Codex & 1.75 & 14.00\\
GPT-5.6-Luna & 0.20 & 1.20\\
GPT-5.6-Terra & 2.00 & 12.00\\
GPT-5.6-Sol & 5.00 & 30.00\\
Grok-4.5 & 2.00 & 6.00\\
MiniMax-M2.1 & 0.30 & 1.20\\
\bottomrule
\end{tabular}
\end{singlespace}
\end{table}

\section{Failure Analysis}
\label{supp:failures}
We examine the final answers and execution traces from 1,000 RStatBench runs in each retrieval condition to assess whether DARE changes the statistical procedures used by RCodingAgent. Incorrect runs decrease from 682 without DARE to 378 with DARE. Table~\ref{tab:failure_manifestations} shows that wrong numerical or statistical results decrease from 358 to 209, although they remain the most frequent failure in both conditions. Among the 1,000 matched task--model pairs, 333 runs change from incorrect to correct with DARE.
\begin{table}[htbp]
\begin{singlespace}
\centering
\small
\caption{Failure manifestations in the final judged answers.}
\label{tab:failure_manifestations}
\begin{tabular}{lrrr}
\toprule
\textbf{Failure manifestation} & \textbf{Without DARE} & \textbf{With DARE} & \textbf{Change} \\
\midrule
Non-completion & 173 & 35 & $-138$ \\
Requested result omitted & 19 & 13 & $-6$ \\
Wrong method, target, or extraction & 5 & 3 & $-2$ \\
Wrong categorical or identifier result & 14 & 5 & $-9$ \\
Wrong count, shape, or selection & 113 & 113 & 0 \\
Wrong numerical or statistical result & 358 & 209 & $-149$ \\
\midrule
\textbf{Total} & \textbf{682} & \textbf{378} & \textbf{$-304$} \\
\bottomrule
\end{tabular}
\end{singlespace}
\end{table}

The paired traces illustrate changes in statistical method selection. In Task 2460, the request specifies a test of equal cumulative distribution functions for two log-concave samples, based on log-concave maximum-likelihood estimates and 999 permutations. Without DARE, DeepSeek-V4-Pro uses a generic Anderson--Darling test and reports $p=0.0015$. With DARE, it uses \texttt{logcondens::logconTwoSample} with the requested statistic and permutation count, reports $p=0.001$, and identifies the log-concavity assumption in its final interpretation.

In Task 4807, the query requests \texttt{minque::lmm.jack} and a confidence interval for the Female variance component. Without DARE, DeepSeek-V4-Pro instead fits \texttt{lme4::lmer} and constructs its own grouped jackknife, reporting $[-136.90,435.12]$. With DARE, it uses \texttt{lmm.jack} with \texttt{JacNum=10}, \texttt{JacRep=1}, and \texttt{ALPHA=0.05}, obtaining $[-34.41489,170.38]$, which matches the reference interval after rounding.

\end{document}